\documentclass[aps,amssymb]{revtex4}
\usepackage{epsfig} 

\usepackage{graphicx}
\usepackage{epstopdf}

\usepackage[dvips, bookmarks, colorlinks=true, plainpages = false, citecolor = red, urlcolor = blue, filecolor = blue]{hyperref}
\usepackage{amsmath,color}

\definecolor{blue}{rgb}{0,0,1}
\definecolor{red}{rgb}{1,0,0}
\definecolor{green}{rgb}{0,1,0}

\definecolor{brillig}{rgb}{1,.6,0}
\definecolor{blarg}{rgb}{.9,.6,.2}

\newcommand{\be}{\begin{equation}}
\newcommand{\ee}{\end{equation}}
\newcommand{\bea}{\begin{eqnarray}}
\newcommand{\eea}{\end{eqnarray}}
\def\nn{\nonumber} 
\newcommand{\ba}[1]{\begin{array}{#1}}
\newcommand{\je}[2]{\mathrm{#1}\left(#2 K'|m\right)}
\newcommand{\ea}{\end{array}}

\def\G{\Gamma}

\newcommand{\drv}[2]{\partial_{#2} #1}

\def\e{\varepsilon} 
\def\k{\kappa}

\def\f{\phi}
\def\y{\psi}

\newcommand{\ci}{\mathrm{i}}

\newcommand{\Imag}{\mathrm{Im}}

\newcommand{\I}{\mathrm{i}}
\newcommand{\IL}{\mathcal{I}_\ell}
\newcommand{\IR}{\mathcal{I}_r}
\newcommand{\AT}{{A}}
\newcommand{\AL}{{A_\ell}}
\newcommand{\AR}{{A_r}}
\newcommand{\AX}{{A_x}}
\newcommand{\BR}{{B_r}}
\newcommand{\BL}{{B_\ell}}
\newcommand{\CR}{{C_r}}
\newcommand{\CL}{{C_\ell}}
\newcommand{\Ho}{{H}}
\newcommand{\Ve}{{V}}
\newcommand{\AV}{{\overline{A}}}
\newcommand{\BT}{{B_t}}
\newcommand{\BB}{{B_b}}
\newcommand{\schem}[2]{\makebox[.03 \columnwidth]{\raisebox{1ex}{$#1$}}
   \includegraphics[width=.15 \columnwidth]{#2}}

\begin{document}

\title{Cluster densities at 2-D critical points in rectangular geometries}

\author{Jacob J. H. Simmons}
\email{simmonsj@uchicago.edu}
\affiliation{James Franck Institute, University of Chicago, 929 E 57th Street, Chicago, IL 60637, USA}

 \author{Peter Kleban}
\email{kleban@maine.edu}
\affiliation{LASST and Department of Physics \& Astronomy,
University of Maine, Orono, ME 04469, USA}

\author{Steven M. Flores}
\email{smflores@umich.edu}
\affiliation{Department of Mathematics, University of Michigan, Ann Arbor, MI 48109-1043, USA}

\author{Robert M. Ziff}
\email{rziff@umich.edu}
\affiliation{Michigan Center for Theoretical Physics and Department of Chemical Engineering, University of Michigan, Ann Arbor MI 48109-2136, USA}

\keywords{Percolation, Crossing Probabilities, Conformal Field Theory.}

\begin{abstract}
Making use of the complete calculation \cite{SimmonsKleban11} of the chiral six-point correlation function
\be \nonumber
C(z) = \langle \phi_{1,2}\phi_{1,2} \Phi_{1/2,0}(z, \bar z) \phi_{1,2}\phi_{1,2} \rangle \; ,
\ee
with the four $\phi_{1,2}$ operators at the corners of an arbitrary rectangle and the point $z = x+ {\rm i}\, y$ in the interior,  for arbitrary central charge (equivalently, SLE parameter $\k > 0$), we calculate various quantities of interest  for percolation ($\k = 6$) and many other two-dimensional critical points.  In particular, we use $C$  to specify the density at $z$ of critical clusters conditioned to touch either or both  vertical sides of the rectangle, with these sides `wired,' i.e.\ constrained to be in a single cluster, and the horizontal sides free.  These quantities probe the structure of various cluster configurations, including those that contribute to the crossing probability.  

We first examine the effects of boundary conditions on $C$ for the critical $\mathrm{O}(n)$ loop models in both high and low density phases and for both Fortuin-Kasteleyn (FK) and spin clusters in the critical $Q$-state Potts models.  A Coulomb gas analysis then allows us to  calculate the cluster densities with various conditionings  in terms of  the conformal blocks calculated in \cite{SimmonsKleban11}.  Explicit formulas generalizing Cardy's  horizontal crossing probability to these models (using previously known results) are also presented. 

These solutions are employed to generalize previous results demonstrating factorization of higher-order correlation functions to the critical systems mentioned.  An explicit formula for the density of critical percolation clusters that cross a rectangle horizontally with free boundary conditions is also given.  Simplifications of the hypergeometric functions in our solutions for various models are presented. 

High precision simulations verify these predictions for percolation and for the $Q=2$ and $3$-state Potts models, including both FK and spin clusters. Our formula for the density of crossing clusters in percolation in open systems is also verified.

\end{abstract}
\maketitle

\section{Introduction}

The conformal invariance of crossing probabilities in critical percolation was first suggested by Aizenman, on the basis of numerical results by Langlands et al.\ \cite{LanglandsEtAl92}.  This prompted Cardy \cite{Cardy92} to apply the methods of boundary conformal field theory (CFT)  to derive his celebrated formula for the horizontal crossing probability in a rectangle of arbitrary aspect ratio. This formula is an excellent example of the predictive power of CFT.  

Recently there has been a renewed interest in crossing probabilities as prototype non-local observables; rigorous proofs have appeared \cite{Smirnov}, and progress has been made in exploiting CFT and other methods  to capture non-local phenomena \cite{MathieuRidout07,Kytola08}.  This includes our previous crossing results for percolation  \cite{SimmonsKlebanZiffJPA07}.  Much of the contemporary research has been motivated by the development of Schramm-Loewner Evolution (SLE) \cite{LawlerSchrammWerner01}, which provides a rigorous framework for  conformal symmetry
but with a  perspective contrasting CFT.   Work on crossing probabilities has also stimulated new research in number theory, in particular the development of the theory of higher-order modular forms.  See \cite{KlebanZagier03} and \cite{DiamantisKleban09} for details and references.

In \cite{KlebanSimmonsZiff06} and \cite{SimmonsKlebanZiffPRE07} we established, by use of conformal field theory and high-precision simulation, exact and universal factorizations of certain higher-order correlation functions in terms of lower-order correlation functions for percolation clusters in two dimensions at the percolation point.  In that work, the correlation functions involved the density of critical percolation clusters constrained to touch one or two isolated boundary points, or single boundary intervals.  The question of which conformal correlation functions factorize in this way is examined in \cite{SimmonsKleban09}.

A more recent paper \cite{SimmonsZiffKleban08} considers percolation densities in a rectangle conditioned to touch one or both vertical sides, with the sides `wired', i.e.\ constrained to belong to a single cluster.  Evaluating these quantities via CFT requires calculation of the correlation function (\ref{cf1})
\be \nn
C(z)=\langle \phi_{1,2}^c(0)\phi_{1,2}^c(\ci) \Phi_{1/2,0}(z, \bar z) \phi_{1,2}^c(R)\phi_{1,2}^c(R+\ci) \rangle_{\cal R}\; ,
\ee
in a rectangular geometry  $\mathcal{R}$ with the boundary operators at the corners.
This is tantamount to the challenging task of evaluating a chiral six-point function in the complex plane.   In \cite{SimmonsKleban11} all solutions of the PDEs governing this correlation function are determined, for arbitrary central charge (or equivalently, SLE parameter $\k > 0$) in terms of algebraic and Appell hypergeometric functions.

The solutions found in \cite{SimmonsKleban11} are employed here to give weights and cluster densities for various quantities in a range of critical models. (We use the term ``cluster density"  in a particular way; the definition is given in subsection \ref{pie}.)      We index the models via the SLE parameter $\k$ in this article, as it includes both branches of the $\mathrm{O}(n)$ model and thus the FK and Potts spin cluster models. The interpretation of $\Phi_{1/2,0}$ as a density operator holds as long as its weight is positive, or when $8/3\le\k\le8$ (corresponding to $0\le n \le 2$).  Outside of this region, we do not offer  a physical interpretation for $C(z)$, though the solutions found in  \cite{SimmonsKleban11} remain valid. 

We also employ the results for $C$ to obtain generalized factorization of correlations.  In  \cite{SimmonsZiffKleban08}, which treats percolation only,  a certain universal ratio of cluster densities within a rectangular region is considered, involving the densities of clusters touching the left and/or right sides of the rectangle with wired, or fixed, boundary conditions on those sides. The ratio  is nearly constant everywhere in the rectangle, varying by less than 3\%, and in addition does not depend on the vertical position $y$ in the rectangle.  In this paper, we generalize the percolation results to the range of critical models mentioned.  How exactly the factorization holds depends on the model, but the independence from $y$ is always valid.

From a broader perspective, $C(z)$ provides, via the densities,   information on the structure of crossing and related cluster configurations (or equivalently loop configurations) in the critical models mentioned.  In this sense it generalizes and deepens Cardy's analysis of crossing in percolation \cite{Cardy92}.

	  In section \ref{Th} we recall the results for the correlation function (\ref{cf1}) obtained in \cite{SimmonsKleban11}. The interesting independence from one coordinate mentioned (see  (\ref{DE3})) appears here.    Subsection \ref{pie} then examines questions that arise in applying our results to the specific critical models mentioned.  In particular, we examine the effects of boundary conditions, and obtain formulas for  properly normalized densities.  In addition, we give explicit formulas for a certain generalization of Cardy's horizontal percolation crossing probability \cite{Cardy92}.   Subsection \ref{Icc} identifies the solutions for the densities of various cluster configurations of interest by use of Coulomb gas methods.  

In section \ref{factor}  we  analyze a certain universal ratio $\rho$ of densities, which (in the scaling limit) is equal to a ratio of involving critical cluster densities and crossing probabilities.  This quantity is universal, and almost constant everywhere in the rectangle, regardless of aspect ratio, in many models, which implies an almost exact factorization of the corresponding correlation functions (or densities).  This work generalizes the ratio examined in \cite{SimmonsZiffKleban08} for percolation.

Section \ref{Dpcc}  returns to the case of percolation.  Here we take advantage of the ``locality" property of percolation, as it is referred to in SLE (equivalent to the independence of local sites or bonds in lattice models) to derive a formula for the density of horizontal crossing clusters $P_{A_x}$ in rectangles with free boundary conditions on all sides.

In section \ref{Sims} we compare our predictions for the ratio $\rho$, which governs factorization, and our formula for the density of percolation clusters crossing a rectangle with high-precision simulations.  The agreement is very good.

Section \ref{Summ} contains a detailed summary of our results.  

The correlation function $C(z)$ determines various densities; a similar correlation function without the $\Phi_{1/2,0}$ determines crossing weights.  These in turn depend on the Appell or $_2F_1$ hypergeometric functions, respectively.   Appendix \ref{Spv} presents results for various critical models where the hypergeometric functions simplify.  

\section{Theory} \label{Th}
\subsection{Results for the correlation function} \label{cfDEs}

In this subsection we recall results obtained in \cite{SimmonsKleban11} for  the six-point correlator  
\be \label{cf1}
C(z)=\langle \phi_{1,2}^c(0)\phi_{1,2}^c(\ci) \Phi_{1/2,0}(z, \bar z) \phi_{1,2}^c(R)\phi_{1,2}^c(R+\ci) \rangle_{\cal R}\; ,
\ee 
in the rectangular geometry  $\mathcal{R}:= \{z=x + \ci y\in\mathbb{C}\, |\, 0<x<R, 0<y<1\}$. The aspect ratio is given by 
\be \label{Rvsm}
R=\frac{K(m)}{K'(m)} \; ,
\ee
where $K'(m):=K(1-m)$, with
 $K$ is the complete elliptic integral.  Conversely, the elliptic parameter $m$ specifies the aspect ratio $R$ via
 \be  \label{mvsR}
m=\frac{\vartheta_4{}^4\left(0,e^{- \pi R} \right)} {\vartheta_3{}^4\left(0,e^{- \pi R} \right)}\; ,
\ee
(Note that $m$  differs from  the standard modular lambda parameter, which is $1-m$ here.)
 
The conformal dimensions and central charge are  
\bea \label{hcdims}
h_{1/2,0} &=& \bar h_{1/2,0}=\frac{(8-\k)(3\k-8)}{64 \k} \\
 h_{1,2} &=& \frac{6-\k}{2\k} \\
 h_{1,3} &=& \frac{8-\k}{\k} \\
 c &=& \frac{(3 \k -8)(6-\k)}{2 \k} \; ,
 \eea
 where $\k$ is the Schramm-Loewner Evolution (SLE) parameter and $ h_{1,3}$ is used below.    Making use of the coordinates 
 \be \label{xipsidef}
\xi :=\je{sn}{x}^2\; , \qquad \mathrm{and} \qquad \y :=\mathrm{sn}\left(y\, K'|1-m\right)^2\; ,
\ee
 we find that any solution that is a single conformal block can be written either in the form  
  \be \label{CeqfG}
C(z) = f(\xi,\psi,m) \, G\left(\xi,m \right) \; ,
\ee 
where the algebraic prefactor $f$ is given by
\be \label{fxipsi}
 f(\xi,\psi,m) = c(m)  \left[ \frac{(1-m \xi^2)^2}{ \xi (1-\xi) (1-m \xi)} + \frac{(1-(1-m) \psi^2)^2}{ \psi (1-\psi) (1-(1-m) \psi)} -4 \right]^{-h_{1,3}/2+h_{1/2,0}} \; ,
\ee
with $c(m)$ given by
\be \label{fconst}
c(m) = 2^{ h_{1,3}}(K')^{8h_{1,2}+2 h_{1/2,0}}(m(1-m))^{2h_{1,2}} \; ,
\ee
or as in (\ref{CeqfG}) with $G\left(\xi,m \right) \to G\left(\y,1-m \right)$.  

The coefficient $ 2^{ h_{1,3}}$ indicates the presence of a corner operator.  The prefactor $f$ is  a $\k$-dependent power of the density of clusters in a rectangle attached to a fixed boundary.  See \cite{SimmonsKleban11} for details.

The algebraic prefactor $f$  is independent of boundary conditions.  The conformal block $G$, on the other hand, is strongly dependent on them, as discussed below.  This plays an important role in our results.  Now $G$ is given by  one of the five solutions  
\begin{align} \label{GI}
G_{\mathrm{I}}(\xi,m)&=\frac{\G(2-8/\k)\G(16/\k -1)}{\G(12/\k)\G(1-4/\k)}\frac{\left[m(1-m)\right]^{2/\k}\xi ^{8/\k-1/2}}{\left[(1-\xi)(1-m \xi)\right]^{4/\k-1/2}}F_1\left(1-\frac{4}{\k };\frac{4}{\k },\frac{4}{\k };\frac{12}{\k }\bigg| \xi ,m \xi \right) \; , \\ \label{GII}
G_{\mathrm{II}}(\xi,m)&=\frac{(1-m)^{2/\k}}{m^{6/\k-1}\left[\xi (1-\xi )(1-m \xi )\right]^{4/\k-1/2}}F_1\left(1-\frac{4}{\k};\frac{4}{\k},2-\frac{16}{\k};2-\frac{8}{\k}\bigg|1-m,1-m \xi \right) \; , \\ \label{GIII}
G_{\mathrm{III}}(\xi,m)&=\frac{(1-m)^{2/\k }}{m^{6/\k-1 }\left[\xi (1-\xi )(1-m \xi )\right]^{4/\k-1/2}}F_1\left(1-\frac{4}{\k};\frac{4}{\k},2-\frac{16}{\k};2-\frac{8}{\k}\bigg| m,m \xi \right) \; , \\ \label{GIV}
G_{\mathrm{IV}}(\xi,m)&=\frac{(1-m \xi )^{12/\k-3/2}}{\left[m(1-m)\right]^{6/\k-1}\left[\xi (1-\xi )\right]^{4/\k-1/2}}F_1\left(1-\frac{4}{\k};\frac{4}{\k},2-\frac{16}{\k};2-\frac{8}{\k}\bigg|1-m,\frac{1-m}{1-m \xi }\right) \; {\rm and}\\ \label{GV}
G_{\mathrm{V}}(\xi,m)&=\frac{\G(2-8/\k)\G(16/\k-1)}{\G(12/\k)\G(1-4/\k)}\frac{m^{2/\k}(1-\xi )^{8/\k-1/2}}{(1-m)^{6/\k-1}\xi ^{4/\k-1/2}(1-m \xi )^{1/2}}F_1\left(1-\frac{4}{\k};\frac{4}{\k},\frac{4}{\k};\frac{12}{\k} \bigg |\frac{m(1-\xi )}{1-m \xi },\frac{1-\xi }{1-m \xi }\right) ,
\end{align}
where $F_1$ is the first Appell function
\be \label{F1def}
F_1\left(a;b_1,b_2;c|z_1,z_2\right):=\sum_{i,j=0}^\infty \frac{(a)_{i+j}(b_1)_i(b_2)_j z_1{}^iz_2{}^j}{i!\, j!\, (c)_{i+j}}
\ee 
with the Pochhammer symbol $(z)_n := \Gamma(z+n)/\Gamma(z)$.

For our ranges of $\xi$ and $m$ values ($0\leq\xi,m\leq1$), (\ref{GI}-\ref{GV}) exhaust the convergent Frobenius  series solutions to the differential equations that can be expressed with a single $F_1$. 
One can also use one of these solutions (or the five mentioned below) with $\{ \xi \rightarrow \y ,\, m   \rightarrow  1-m \}$.  Each of these is a single conformal block.  Only three of them are independent, as discussed in section II A of \cite{SimmonsKleban11}.

There are also five other convergent solutions that can be expressed with a single second Appell function $F_2$ (see \cite{SimmonsKleban11} for details), of which we use only one here:
\bea \label{GVI}
G_{\rm VI}(\xi,m)&:=&G_{\rm II}-n G_{\rm I} =  G_{\rm IV}-nG_{\rm V} \\ \nn
&=&\frac{\G(2-8/\k)\G(16/\k-1)\G(4/\k)}{\G(1-4/\k)\G(8/\k)^2}[m(1-m)]^{2/\k}\left[\frac{\xi(1-\xi)}{1-m \xi}\right]^{8/\k-1/2}F_2\left(\frac{16}{\k}-1;\frac{4}{\k },\frac{4}{\k };\frac{8}{\k },\frac{8}{\k }\bigg| 1-\xi,\frac{\xi(1-m)}{1-m \xi} \right)\; ,
\eea
where $n$ is  the parameter of the $\mathrm{O}(n)$ loop models, given by
\be \label{loopeq}
n=-2 \cos(4 \pi/\k) \; ,
\ee 
with $F_2$ defined by
\be \label{F2def}
F_2\left(a;b_1,b_2;c_1,c_2 | z_1,z_2\right):=\sum_{i,j=0}^\infty \frac{(a)_{i+j}(b_1)_i(b_2)_j z_1{}^iz_2{}^j}{i!\, j!\, (c_1)_i (c_2)_j}\; .
\ee 

The fact that these solutions for $G$ only depend on two variables rather than three follows on combining the conformal PDEs.  One finds that
\be \label{DE3}
\drv{\drv{G}{\xi}}{\y}(\xi, \y, m)=0\;.
\ee
This equation is the basis for the $y$-independence of the factorization behavior discussed below.  It indicates the presence of an unknown symmetry.

For use below, we also recall the behavior of $\xi$, $\y$ and  $m$ under the symmetries of the rectangle: mirror symmetries about $x=R/2$, $y=1/2$ and $x=y$.  These symmetry operations translate, respectively, into 
\be  \label{Sym}
( \xi, \y, m ) \to  \left( \frac{1-\xi}{1-m\, \xi}, \y, m\right), \left(\xi, \frac{1-\y}{1-(1-m) \y}, m \right), \left(\y, \xi,1-m \right) \; .
\ee


\subsection{Physical interpretation of equation (\ref{cf1}) and boundary conditions} \label{pie}

The correlation function (\ref{cf1}) describes cluster densities for a variety of critical statistical mechanics models; in this section we examine this correspondence. We focus in particular on the Potts models and their boundary conditions in order to obtain explicit formulas.  These results allow us to implement simulations (see section \ref{Sims}) to test our predictions.  We begin with critical ${\mathrm O}(n)$ loop models (in the continuum limit) because they furnish a clear picture for these densities, and describe a continuum of critical points with $0\leq n\leq2.$ There are many realizations of these models but in each case the degrees of freedom are closed fractal loops, each contributing to the partition function with a fugacity $n$. There are two critical branches depending on the energy cost per loop length, representing dilute and dense loop phases.
As noted, we use the SLE parameter $\kappa$ to index our models. This is convenient as $\kappa$ nicely parameterizes both critical branches of the ${\mathrm O}(n)$ loop model; the two parameters are related by (\ref{loopeq})
with $\kappa\leq4$ (resp. $\kappa > 4$) corresponding to the dilute (resp. dense) phase.

We can condition the loop ensembles so that open loop segments emerge from pairs of points on the boundary.  We call these loops Ôboundary arcsÕ  to distinguish them from the closed loops in the bulk. These boundary arcs are equivalent to SLE traces, and in CFT are implemented by placing $\phi_{1,2}$ Kac operators at their endpoints on the boundary. Because they are distinguished from bulk loops, we can give the boundary arcs a fixed weight of $1$. The correlation functions we consider correspond to an ${\mathrm O}(n)$ model with boundary arcs attached to the four corners of an arbitrary rectangle.

In the ${\mathrm O}(n)$ model, bulk points are said to be {\it adjacent} to a given boundary segment whenever a path can be drawn from the point to the segment without crossing any loops or boundary arcs. It has been argued using Coulomb gas techniques that the density operator for these adjacent points is $\Phi_{1/2,0}(z,\bar{z})$ \cite{NienhuisRiedelSchick_1980}.  Mathematically our conformal block expressions ((\ref{GI})-(\ref{GV}) and (\ref{GVI})) are valid for all $\k > 0$, but we only discuss physical interpretations for $8/3 \leq \k \leq8$. Outside of this range the density operator in the ${\mathrm O}(n)$ models has weight zero, while $\Phi_{1/2,0}$ does not. 

In order to proceed, we also need to consider generalizations of Cardy's crossing probability.  They provide an example of applying boundary conditions appropriate to the Potts models, and a necessary normalization.  We discuss these matters in subsections \ref{xingprob} and \ref{FKformulas} below.

In applying our results to model systems, three kinds of quantities arise.  Unnormalized weights, which are denoted by $\Pi$, crossing probabilities denoted by $P$, and cluster densities, also denoted by $P$. The crossing probabilities are true probabilities, and can be completely determined from our CFT results.  For a given model, they depend on the aspect ratio $R$ only.  A cluster density (or more simply, ``density") is a bit more complicated.   By ``density" we understand  the scaled probability in a small region around a point with some specified conditioning, e.g.\ connected to a specific boundary segment. This is the probability, on a lattice with mesh size $\delta$, that a neighborhood of $z$ with characteristic length $\epsilon$ includes points with the specified conditioning,  divided by the scaling factor $\epsilon^{2h_{1/2,0}}$ and taken in the limit $\delta\ll\epsilon\rightarrow0$. These densities are proportional to the correlation function (\ref{cf1}), divided by a partition function that for a given model is  a four-point function of $\phi_{1,2}$ operators. (The partition function depends on $R$ but not $z$.)  This determines the density up to a non-universal factor that depends on the details of how the $\Phi_{1/2,0}$ operator in (\ref{cf1}) is regularized.  This factor depends on the particular model and is thus not given by CFT.  We will ignore it in the formulas below.   However, in the case of a ratio such as $\rho$ (see section \ref{factor}) the nonuniversal factors divide out, so the conformal result is complete.

\subsubsection{Crossing Probabilities} \label{xingprob}

\begin{figure}[htbp] 
  \centering
  \schem{\Ho}{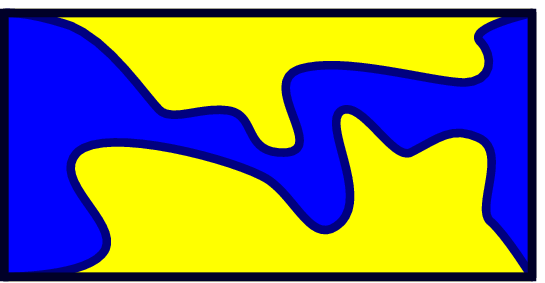}
  \hspace{.05 \columnwidth}
    \schem{\Ve}{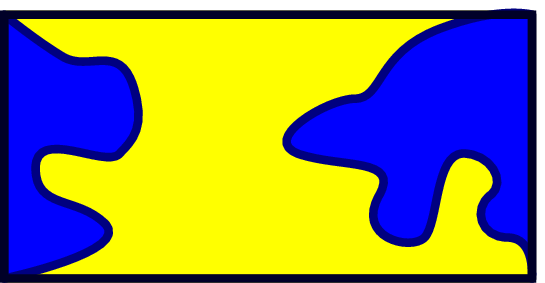}
  \caption{Configurations for the generalization of Cardy's formula.  \label{crdy}}
\end{figure}

We consider the crossing probabilities for the ${\rm O}(n)$ model in a rectangle with boundary arcs attached to the corners.
Inside the rectangle, the boundary arcs join in one of two ways: they may connect the two top and two bottom corners, which we call configuration $H$, or they may connect the two left and two right corners, which we call configuration $V$ as shown in figure \ref{crdy}.    We use $\Pi_H$ (resp.\ $\Pi_V$) to denote the weight of the corresponding configurations  for the ${\mathrm O}(n)$ model, specifically when the boundary arcs have weight $1$.  

We now consider the dense ${\mathrm O}(n)$ model phase. Recall that the Potts model can be exactly mapped onto a representation known as the Fortuin-Kastelyn (FK) random cluster model, which replaces the sum over spin configurations with a sum over nearest neighbor bond configurations \cite{FortuinKasteleyn1972}. This is the basis of the Swendsen-Wang algorithm frequently used to simulate the Potts model \cite{sw}.   At the critical point, replacing the sum over bond configurations in the FK model partition function with a sum over loops following the cluster boundaries gives a mapping to the dense phase of the ${\mathrm O}(n)$ model, where each loop contributes a weight $n=\sqrt{Q}$ to the partition function.

 In the FK representation, the $\phi_{1,2}$ operators that mark the ends of SLE traces are equivalent to changes between free and wired boundary conditions. On a `free' boundary there is no conditioning and the bond configurations are summed over freely as in the bulk, while on a `wired' boundary  all bonds are occupied, so that the boundary sites all belong to the same FK cluster.  Thus there are two possible interpretations for the correlation function (\ref{4opeq}): either the pairs of operators $\phi_{1,2}(0)\phi_{1,2}({\rm i})$ and $\phi_{1,2}(R)\phi_{1,2}(R+{\rm i})$ represent the left and right wired sides of the rectangle respectively, or $\phi_{1,2}(0)\phi_{1,2}(R)$ and $\phi_{1,2}({\rm i})\phi_{1,2}(R+{\rm i})$ represent wired bottom and top sides. 
We will adopt the former interpretation (unless explicitly stated otherwise), so the SLE curves follow the boundaries of the FK clusters anchored to the left and/or right sides. 

For ${\rm O}(n)$ model configurations of type $V$, the boundary arcs close into two loops that surround the clusters connected the left and right sides.  Thus the contribution of each such configuration to the FK partition function gains a factor $n^2$ and one has
\be \label{ZV}
Z_V = n^2 \, \Pi_V \; .
\ee
For configurations of type $H$, the boundary arcs close into a single loop around the crossing cluster.  Thus the contribution to the  FK  partition function gains a single factor of $n$ and
\be \label{ZH}
Z_H = n \, \Pi_H \; .
\ee
The total FK partition function is therefore
\be \label{ZFK}
Z =  n \, \Pi_H +  n^2 \, \Pi_V \; .
\ee
The corresponding result when the top and bottom sides are wired follows on interchanging $H$ and $V$.

At this point we notice a subtlety that will become important when we  consider spin cluster densities, namely whether the fixed spins  on the right side match the fixed spins on the left side.
We next construct an FK model partition function $\bar Z$ that only samples configurations where the left and right sides are \emph{mutually fixed} to the same spin.  This is achieved by adding, to all configurations, a single occupied bond  between the left and right hand wired sides.  Thus, the left and right sides belong to the same cluster, so they are always wired to the same spin.  Now for the ${\rm O}(n)$ model, the boundary arcs must follow the new bond between the left and right edges, so the boundary arcs close along the top and bottom sides.  The consequences for the FK partition function are as follows.  For type $V$ configurations, the boundary arcs close into a single loop.  Thus
\be \label{ZVextra}
{\tilde Z}_V = n \, \Pi_V \; ,
\ee
where ${\tilde Z}$ indicates an FK partition function with the extra occupied bond included.  For type $H$ configurations, the left and right side are already connected, so the boundary arcs close into two loops.  They can be envisaged by placing the new bond above the top edge.  Then  one extra loop follows the top  edge of the horizontal crossing cluster but is below the new bond, and the other follows the bottom edge of the horizontal crossing cluster, and goes around both left and right sides and above the extra bond.  Thus 
\be \label{ZHextra}
{\tilde Z}_H = n^2 \, \Pi_H \; .
\ee
The net result for the extended FK partition function is therefore
\be \label{ZFKextra}
{\tilde Z} =  n^2 \, \Pi_H +  n \, \Pi_V \; .
\ee
Note that the forms $Z$ and $\tilde Z$ are symmetric under exchanges of the labels $H$ and $V$.  This is a consequence of the  duality transformation that maps the rectangle with left and right sides wired and an extra bond onto a rectangle with top and bottom wired and a similar extra bond that is dual to the original bond.

Now the rectangle with an extra bond is not physical. Thus we need to adjust ${\tilde Z}$.  The necessary factor is determined from the explicit form of the critical FK partition function, where each occupied bond contributes a factor of $n$.   Adding an extra occupied bond multiplies each term in the partition function by an extra factor of $n$.  Dividing $\tilde Z$ by $n$, we find the FK partition function for mutually fixed left and right sides:
\be \label{ZFKmut}
\bar Z =  n \, \Pi_H +   \, \Pi_V \; .
\ee
As above, the corresponding result when the top and bottom sides are mutually wired follows on interchanging $H$ and $V$.

The result (\ref{ZFKmut}) is what we expect.   Comparing (\ref{ZFK}) and (\ref{ZFKmut}) we see that the weight of an $H$ configuration is insensitive to mutual wiring, while the  $V$ weights loose a factor of $Q=n^2$.

We now consider the dilute phase of the ${\rm O}(n)$ model.  This can be related to spin clusters in the $Q$-state Potts model. In contrast to the FK representation, this association does not follow from direct manipulation of the Potts and ${\mathrm O}(n)$ models. Instead one invokes a result known as SLE duality \cite{RohdeSchramm2005}, which states that for $\kappa>4$ the outer hull of an SLE with $\kappa$ has the same fractal dimension as an SLE with $\kappa'=16/\kappa = (4/\pi)\arccos(-\sqrt{Q}/2)$, and furthermore that these related values of $\kappa$ correspond to CFTs with the same central charge. If the spin clusters can be described by an SLE process then it must have parameter $\kappa'$.

The boundaries of spin clusters separate regions with a single spin value, say $s = 1$, from neighboring regions with $s\neq1$.   An SLE that describes these domain walls should separate boundary regions with $s=1$ and $s \neq1$ too. For the Ising model, the $\kappa = 3$ SLE trace represents a change between $+$ and $-$ boundary conditions \cite{ChelkakSmirnov2009}.
For the 3-state Potts model it has been convincingly argued that the change between $s = 1$ and $s = 2$ or 3 (boundary spins freely summed over 2 and 3) corresponds to an SLE with $\kappa = 10/3$ \cite{GamsaCardy2007}.

In section \ref{Sims} we simulate these dilute phases in a rectangle with $s=1$ boundary conditions on the left and right sides, freely summing over $s\neq1$ spins on the top and bottom edges.  Unlike the FK cluster model, there is only one way to fix the boundaries, by fixing both edges to $s=1$.  This can be implemented by adding an extra nearest neighbor bond between the left and right sides exactly as with the FK clusters; the factors found for the mutually fixed FK partition function carry through in exactly the same form.  Because there is no exact mapping from the spin cluster representation to the dilute ${\mathrm O}(n)$ model we cannot  fix their relative normalization.  Instead, we adopt the $\bar Z$ normalization for the spin clusters.

From the partition functions (\ref{ZFK}) and (\ref{ZFKmut}), we can calculate the probability $P_H$ (resp.\ ${\bar P}_H$)  of a horizontal   crossing  FK (resp.\ either FK or spin) cluster by considering a rectangle with independently (resp.\ mutually) wired left and right sides.    This is simply the ratio of the weight of $H$  configurations with the chosen boundary conditions to the appropriate partition function.  From the above, the weight of $H$ configurations  is $n\Pi_\Ho$ for independently or mutually wired sides, and the weight of $V$ configurations is $n^2\Pi_\Ve$ (resp.\ $\Pi_\Ve$) for independently (mutually) wired sides.  Thus
\begin{align} \label{Hxings} 
 P_H = \frac{\Pi_H}{\Pi_H+n\Pi_V}, && {\bar P}_H = \frac{n\Pi_H}{n\Pi_H+\Pi_V} \; .    
\end{align}
For explicit formulas for $\Pi_H$ and $\Pi_V$ (which make use of results from \cite{BauerBernardKytola05})  see (\ref{Cardy11}), (\ref{Cardy13}), (\ref{G13}) and (\ref{PiVeq}). A result for crossing on spin clusters for the Ising model on a circle, which is conformally equivalent to ${\bar P}_H$,  was obtained in \cite{ArguinSaintAubin2002}.  Note that there are other generalizations of Cardy's formula using different boundary conditions (see  \cite{LawlerSchrammWerner01, RohdeSchramm2005}).

 For $n=1$, which is realized by  percolation (dense phase) or Ising spins (dilute phase), the distinction between mutually wired and independently wired boundary conditions vanishes.  For percolation, the partition function for either type of boundary condition is constant (i.e.\ independent of the rectangle aspect ratio) and is renormalized to 1 in the continuum limit.  Thus, $\Pi_H$ and $\Pi_V$ become Cardy's formula \cite{Cardy92} for horizontal and vertical crossings respectively, with the property $\Pi_H+\Pi_V=1$.

\subsubsection{Cluster densities}  \label{FKformulas}

\begin{figure}[htbp] 
  \centering
  \schem{\AT}{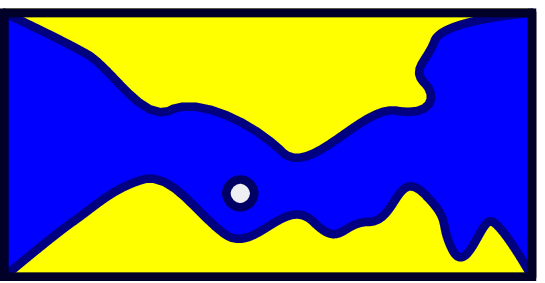}
  \hspace{.05 \columnwidth}
  \schem{\BR}{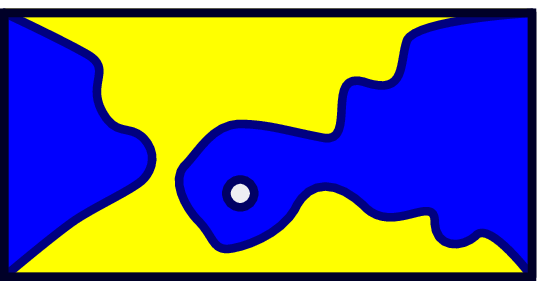}
  \hspace{.05 \columnwidth}
  \schem{\BL}{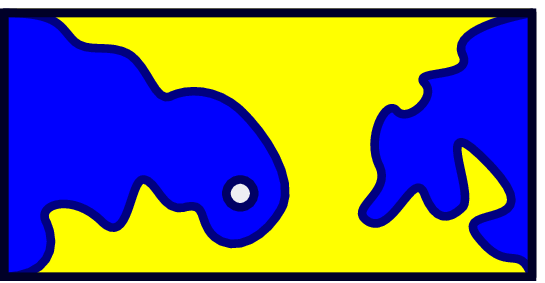}\\[3em]
  \schem{\AV}{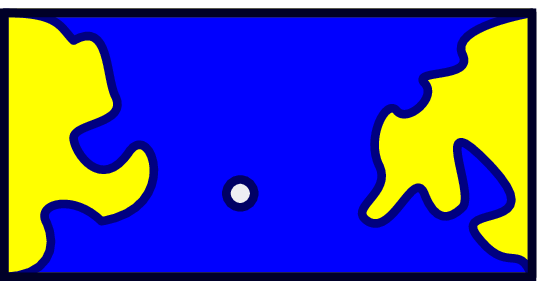}
  \hspace{.05 \columnwidth}
  \schem{\BT}{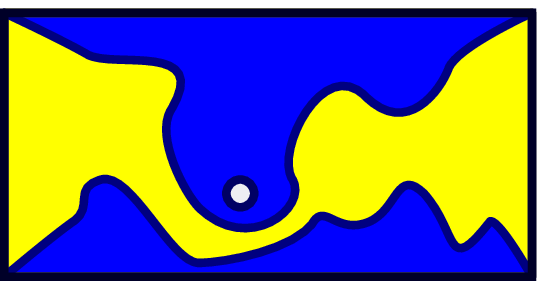}
  \hspace{.05 \columnwidth}
  \schem{\BB}{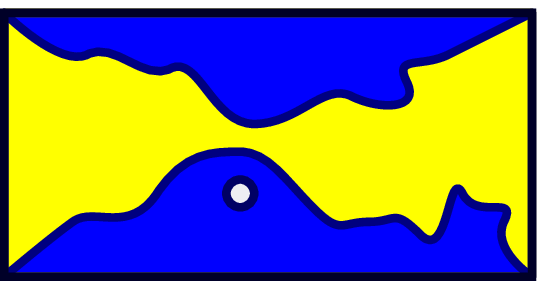}
  \caption{Schematic drawings of the six configurations associated with (\ref{cf1}). We show the boundary arcs and the point $z$, but suppress bulk loops for clarity.  Note that $z$ must be adjacent to the appropriate side(s) of the rectangle.
   \label{config3}}
\end{figure}

The correlation function (\ref{cf1}) involves a density operator at the point $z$, and has contributions from six types of configurations $\{\AT, \BL, \BR\}$ and $\{\AV, \BB, \BT\}$ that depend on the location of $z$ relative the the boundary arcs, as illustrated in figure \ref{config3}.
We now apply our results to find formulas for the cluster densities in the FK and spin cluster Potts models in terms of the weights $\Pi_\Ho$,  $\Pi_\Ve$,  $\Pi_\AT$, $\Pi_{\BR}$ and $\Pi_{\BL}$.  These weights are defined   for the ${\mathrm O}(n)$ model, specifically when the boundary arcs have weight $1$, as for $\Pi_H$ and $\Pi_V$. In the FK cluster case we find the density of sites that belong to the same FK bond cluster as the specified wired edge.  In the spin cluster case we find the density of $s=1$ sites that belong to clusters attached to the specified $s=1$ spin edges.

We now calculate the density $P_\ell(z)$ (resp.\ $P_r(z),P_{\ell r}(z)$) of clusters anchored to the left (resp.\ right, left and right) side(s) of a rectangle with either independently or mutually wired left and right sides.  This is equal to some linear combination of the weights $\{\Pi_A,\Pi_{B_\ell}\}$ (resp.\ $\{\Pi_A,\Pi_{B_r}\}$, $\{\Pi_A\}$) divided by the partition function  (\ref{ZFK}) or (\ref{ZFKmut}).  The coefficients of this linear combination are determined by the boundary conditions in the same way as above.  The boundary arcs of $A$ (resp.\ $B_\ell,B_r$) connect as in $H$ (resp.\ $V$), so the boundary conditions introduce the same factors for $\Pi_A$ as they did for $\Pi_H$ (resp.~the same for $\Pi_{B_\ell},\Pi_{B_r}$, as for $\Pi_V$).  Therefore
\begin{align} \label{Pllr}
P_\ell(z)=\frac{\Pi_A+n\Pi_{B_\ell}}{\Pi_H+n\Pi_V},&&{\bar P}_\ell(z)=\frac{n\Pi_A+\Pi_{B_\ell}}{n\Pi_H+\Pi_V},\\  \label{Prlr}
P_r(z)=\frac{\Pi_A+n\Pi_{B_r}}{\Pi_H+n\Pi_V},&&{\bar P}_r(z)=\frac{n\Pi_A+\Pi_{B_r}}{n\Pi_H+\Pi_V},\\  \label{Plrlr}
P_{\ell r}(z)=\frac{\Pi_A}{\Pi_H+n\Pi_V},&&{\bar P}_{\ell r}(z)=\frac{n\Pi_A}{n\Pi_H+\Pi_V} \; .
\end{align}
As before, the $P$ expressions apply to  the independently wired FK configurations, and the $\bar P$ expressions apply to both the mutually wired FK configurations and the spin cluster configurations.

We could also assume that the top and bottom sides are wired, which would result in the same expressions with $\{\AT, \ell,r,\Ve, \Ho\}$ replaced with $\{\AV, b, t, \Ho, \Ve\}$, reflecting the $(\xi,\, \psi \, m)\rightarrow(\psi, \, \xi, \, 1-m)$ symmetry noted in section \ref{cfDEs}.

 In section \ref{Sims} we compare the densities calculated in this section  with simulations that measure $P_\ell(z),\, P_r(z),\, P_{\ell r}(z)$ for $Q=1$, $2$  and $3$-state FK clusters and $Q=2, \, 3$-state spin clusters.

 In the next section, we identify the conformal blocks of the six-point function (\ref{cf1}) that contribute to the weights $\Pi_\AT$, $\Pi_{\BR}$ and $\Pi_{\BL}$ (see (\ref{pia})-(\ref{pib2})).  For explicit formulas  for $\Pi_H$ and $\Pi_V$ see (\ref{Cardy11}), (\ref{Cardy13}), (\ref{G13}) and (\ref{PiVeq}).


\subsection{Identifying cluster configurations} \label{Icc}
From the solutions to  the PDEs that govern the correlation function (\ref{cf1}) and  the prefactor $f$ (\ref{fxipsi}), to go further we must identify how the weights $\Pi_\AT$, $\Pi_{\BR}$ and $\Pi_{\BL}$  contribute to the various conformal blocks $G$.  This can be done in an elegant way using Coulomb gas vertex operators, which we will now briefly review; for a more complete treatment see, for example, \cite{DotsenkoFateev84, BYB}. %

The Coulomb gas representation makes use of a chiral bosonic variable $\varphi(z,\bar z)$ with action $S=S_O+S_C+S_D$:
$$
S_O = \frac{g}{4 \pi}\int (\nabla \varphi)^2 \mathrm{d}^2x,
\quad
S_C = \frac{\I \alpha_0}{8 \pi} \int \mathcal{R}\, \varphi\, \mathrm{d}^2x,
\quad
S_D = a \int \cos 2 \varphi\, \mathrm{d}^2x\; .
$$
Starting from the free boson action $S_O$, adding the complex term $S_C$ that couples the field to the scalar curvature $\mathcal{R}$ modifies the stress tensor.  This reduces the central charge of the theory from $c=1$ to $c=1-24 \alpha_0{}^2$.
 The field takes discrete values, $\varphi \in \pi \mathbb{Z}$ almost everywhere, due to the term $S_D$. This suppresses any rescaling of the parameter $g$.  At short distances, the boundaries between these discrete values naturally form a loop ensemble, but at long distances, the behavior of the system should still be dominated by $S_O$.   Reconciling these two scales is only possible if $S_D$ is a marginal perturbation, and this condition lets us relate $g$ and $\alpha_0$, leaving a single free parameter.  Thus
\be \label{CGCC}
c=1-24 \alpha_0{}^2=\frac{(6-\k)(3\k-8)}{2 \k}\; ,
\ee
gives $g=4/\k$ and $2 \alpha_0=(1-g) g^{-1/2}$.

Ignoring a subtlety of the zero mode, which is unimportant here, we decompose the boson into holomorphic and antiholomorphic components $\varphi(z, \bar z)=\varphi(z)+\overline \varphi(\bar z)$ in order to write chiral vertex operators,
$$
\mathcal{V}_{\alpha}(z) = \mathrm{e}^{\mathrm{i} \sqrt{2} \alpha \varphi(z)}\; .
$$
Chiral operators are sufficient because our problem includes a boundary.  The energy flux leaving the system is zero, which means $\varphi(z)$ and $\overline \varphi(\bar z)$ are not independent on the boundary and, by analytic continuation, everywhere else.   Specifically, in the upper half plane $\overline \varphi$ is the analytic continuation of $\varphi$ from the lower half plane \cite{Cardy84}.

The parameter $\alpha$ is called the charge of the vertex operator, by analogy with two dimensional electrostatics.
Non-trivial correlation functions must obey a charge neutrality condition.  Since $S_C$ effectively adds a non-local background charge $-2\alpha_0$ to the system the total vertex charges must satisfy $\sum_i \alpha_i = 2\alpha_0$.
The form of charge neutral correlation functions is still determined by the original gaussian action $S_O$,
$$
\langle \prod_i \mathcal{V}_{\alpha_i}(z_i)\rangle =\prod_{j>i} (z_j-z_i)^{2\alpha_j \alpha_i}\; ,
$$
and the  charge neutrality condition is entirely responsible for the differences between the Coulomb gas and free boson correlation functions.

The form of the two point function then implies that a given scaling operator has two possible vertex representations with charges, $\alpha$ or $2 \alpha_0-\alpha$, %
 related to the conformal weight by
$$
h=\alpha(\alpha-2 \alpha_0)\; .
$$

The condition of charge neutrality severely limits the set of computable correlation functions. %
However, screening operators allow one to extend this set of correlations by adding charge without changing the conformal properties of the system.  Screening operators are non-local with zero weight, and are formed by integrating weight one vertex operators around the singular points of the correlator in a non-trivial way. There are two types of screening operators:
$$
Q_{\pm} = \oint \mathrm{d}z \mathcal{V}_{\alpha_{\pm}}(z).
$$
The charges $\alpha_+$ and $\alpha_-$ are the positive and negative solutions of $1 = \alpha(\alpha-2 \alpha_0)\;$ respectively.  Using (\ref{CGCC}) these are
\be
\alpha_+=\sqrt{\frac{\k}{4}}\qquad \mathrm{and} \qquad \alpha_- = -\sqrt{\frac{4}{\k}} \; .
\ee
We parameterize all other charges as
\be
\alpha_{r,s}^{\pm}=\frac{1\pm r}{2} \alpha_+ + \frac{1\pm s}{2} \alpha_-\; .
\ee
When $r,s \in \mathbb{Z^+}$ the charge corresponds to a Kac operator $\phi_{r,s}$.  We can generalize to $r,s \in \mathbb{Z}/2$, so that the density operator has fixed $r$ and $s$ independent of the model parameter $\k$.

\begin{figure}[htbp] 
  \centering
  \includegraphics[width=0.75 \columnwidth]{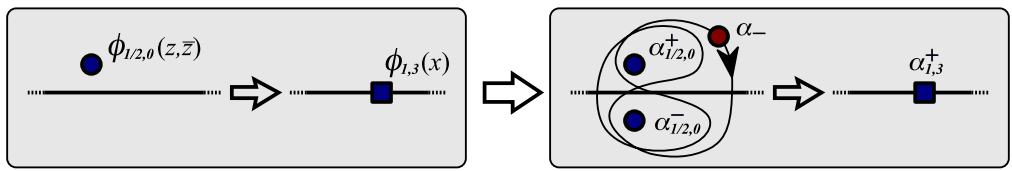}
  \caption{\label{Vertex1} The fusion of a bulk magnetization operator to a free boundary has a unique vertex/screening operator representation.}
\end{figure}

Now we describe how the the vertex operator method helps us to identify the configurations contributing to various conformal blocks.
The charge neutrality condition for a correlation function with two chiral $\phi_{1/2,0}$ and four $\phi_{1,2}$  %
operators allows a unique set of charges and screening operators:
 one $\alpha_{1/2,0}^+$, one $\alpha_{1/2,0}^-$, four $\alpha_{1,2}^-$s and two $Q_-$s.
 Given (\ref{DE3}) we can calculate functions such as (\ref{GI}-\ref{GV}) with the method used in \cite{SimmonsZiffKleban08}, which exploits the common $y$ dependence (assumed in \cite{SimmonsZiffKleban08}) by moving the density operator to the lower boundary, and replacing it with its limiting boundary operator. Thus the bulk $\Phi_{1/2,0}$ operator  becomes  a boundary magnetization operator (see figure \ref{Vertex1}) and we are left with a five point boundary correlation function made up of one $\alpha_{1,3}^+$, four $\alpha_{1,2}^-$s and one $Q_-$.

We choose integration paths for the $Q_-$ that entwine neighboring pairs of boundary operators because, with an appropriately chosen phase, they yield real solutions for arbitrary values of aspect ratio and model parameter.

\begin{figure}[htbp] 
 \makebox[0.05\columnwidth]{\raisebox{1.1cm}{\sf A\,)}}
 \includegraphics[scale=.75]{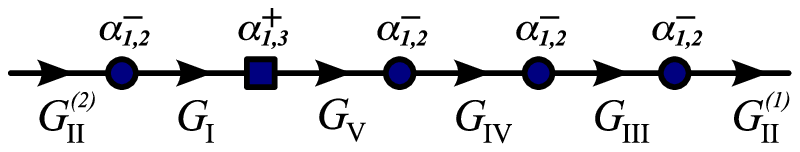}\\
 \makebox[0.05\columnwidth]{\raisebox{1.1cm}{\sf B\,)}}
 \includegraphics[scale=.75]{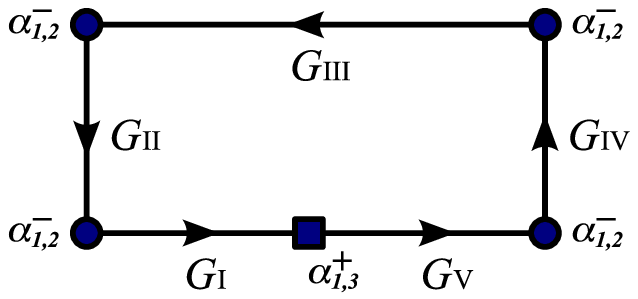}
\hspace{1cm}
 \makebox[0.05\columnwidth]{\raisebox{1.1cm}{\sf C\,)}}
 \includegraphics[scale=.75]{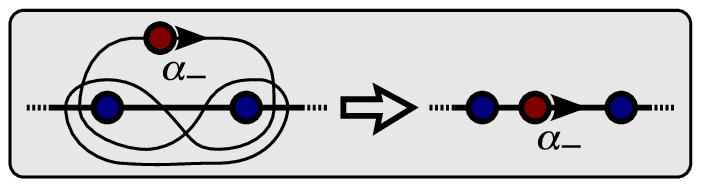}
\caption{ \label{Vertex2} The boundary five point function requires one $\alpha_-$ screening charge.  Integrating the screening operator along the boundary as shown gives the five solutions (\ref{GI}-\ref{GV}).}
\end{figure}

In general we integrate along non-contractible closed paths such as the one on the left hand side of figure \ref{Vertex2}{\sf C}.  These paths can be deformed to run along the real axis except for small circles surrounding the operators.  If $4<\k$ the contribution to the integral from the small circles vanishes with the circle's radius, so we can replace the path with its component along the real axis.%
%
Choosing the paths as in figure \ref{Vertex2}{\sf A} reproduces the indicated conformal blocks from (\ref{GI}--\ref{GV}). The integration is defined in the upper half-plane, and must be transformed into the rectangle as shown in figure \ref{Vertex2}{\sf B}.

For a given correlation function, a conformal block is determined in part by specifying an order in which to fuse the operators, retaining a single conformal family from the OPE under each fusion.  Mathematically, conformal blocks are multiple Frobenius series in small parameters implied by the fusion order with the leading exponents determined by the specified conformal families.  In the vertex operator formalism pairs of vertices that are not entwined with distant vertices have a definite charge when fused together. Such a pair of operators may or may not be entwined by its own screening operator circulating around just the pair. ÊIn either case, such pairs fuse to a single conformal family, which is easily identified by the sum of the fused charges and any entwining screening charges.  Thus a given set of screening charge integration paths in an arbitrary correlation function is simply identified with a conformal block.

Next, as was done in \cite{SimmonsZiffKleban08} for percolation, we use the vertex operator formulation to determine the contribution of the conformal blocks to the configurations of interest.  First,
for the integration paths in figure \ref{Vertex2} we deduce the fusion structure of each block to determine the contributions from the three configurations illustrated in the first row of figure \ref{config3}.  Recall that these weights are defined for the ${\mathrm O}(n)$ models, with  the weight of the boundary  arcs being $1$.  These conformal blocks can be identified with sets of boundary conditions similar to those in section \ref{pie}, except that we may leave unattached arcs instead of  boundary loops.

First consider the block $G_{\mathrm{I}}$ in the limit where the magnetization operator goes to the lower left hand corner.  This  traps the screening operator as well, for a total charge of $\alpha_{1,2}^- + \alpha_{1,3}^+ + \alpha_- = \alpha_{1,4}^+$, which is the charge for a three leg operator.  An $n$-leg operator, associated with $\phi_{1,1+n}$, represents a point where at least $n$ non-contractible boundary arcs (arcs that cannot be smoothly deformed to a point because of some constraint on their behavior) each with weight $1$ are attached to the boundary.
 The only configuration with three or more non-contractable boundary arcs emanating from this corner in that limit is $\BR$; therefore this configuration must be the sole contribution and $G_\mathrm{I} \sim \Pi_{\BR}$.
(Note that this proportionality disregards all $y$ dependence, which is contained entirely in the prefactor $f(\xi,\psi,m)$.)

The block $G_{\mathrm{II}}$ is more subtle.  Fusing the right hand corners together yields a charge of $2 \alpha_{1,2}^-=\alpha_{1,3}^-$, implying a two-leg operator; this excludes $\BL$ configurations which have contractable boundary arcs attached to the right side.

 The remaining contributions to $G_{\mathrm{II}}$ are fixed by considering the left hand edge, which has a net charge of $2 \alpha_{1,2}^-+\alpha_-=0$ indicating that the one leg operators at the corners fuse to the identity family.  If the arcs attached to this edge are non-contractable (as with $\AT$ configurations)  then they must be attached to other $1$-leg operators.  These operators will fix the arc weight at $1$, so the total weight does not change.  If the attached arc is contractable, as with $\BR$ configurations, then bringing the operators together would form a loop with weight $1$, and  a compensating factor of $n$ is needed.  The net result is that $G_\mathrm{II} \sim \Pi_\AT+n\Pi_{\BR}$.
The same result follows by adding a boundary arc between each pair of one-leg operators that fuse to the identity family.

When we consider the block $G_{\mathrm{III}}$ the top edge has net charge $2 \alpha_{1,2}^-+\alpha_-=0$ and fusion is through the identity.  Since there are no other restrictive fusions this block contains contributions from all three configuration types.  The top edge of type $\AT$ configurations has a contractable loop, while $\BR$ and $\BL$ have non-contractible loops attached to the top edge, thus $G_{\mathrm{III}} \sim n \Pi_\AT+\Pi_\BR+\Pi_\BL$.
Again this result can be obtained via  a boundary arc, this time connecting the top corners.  

The associations for $G_{\mathrm{IV}}$ and $G_{\mathrm{V}}$ follow from mirror symmetry and for $G_{\rm VI}$ by definition.  The complete list %
 is %
\bea \label{GI2}
G_\mathrm{I} &\sim& \Pi_\BR\; ,\\ \label{GII2}
G_\mathrm{II} &\sim& \Pi_\AT+n\Pi_\BR\; ,\\ \label{GIII2}
G_{\mathrm{III}} &\sim& n \Pi_\AT+\Pi_\BR+\Pi_\BL\; ,\\ \label{GIV2}
G_\mathrm{IV} &\sim& \Pi_\AT+n\Pi_\BL\; ,\\ \label{GV2}
G_\mathrm{V} &\sim& \Pi_\BL\; , \\ \label{GVI2}
G_{\rm VI} &\sim& \Pi_\AT\; .
\eea

The relative normalization in (\ref{GI}--\ref{GV}) emerges naturally in the vertex operator formalism, since the integral expressions for the different conformal blocks have different integration paths but the same integrand.  Hence the proportionalities in equations (\ref{GI2}--\ref{GVI2}) all omit the same factor.  This is an advantage of using vertex operator over the differential equation analysis of \cite{SimmonsKleban11}, which  would require comparing several limiting cases to fix the relative normalizations.

Using the expression (\ref{fxipsi}) for the common pre-factor $f(\xi,\psi,m)$  we thus find explicit expressions for the weights of the three configuration types $\AT$, $\BR$ and $\BL$, as
\bea \label{pia}
\Pi_\AT  &=& f(\xi,\psi,m)G_{\rm VI}(\xi,m)\\
\Pi_{\BR}&=& f(\xi,\psi,m)\, G_{\rm I}(\xi,m) \qquad\qquad \mathrm{and}\\ \label{pib2}
\Pi_{\BL}&=& f(\xi,\psi,m)\, G_{\rm V}(\xi,m)\; .
\eea
Explicit formulas for $f$ are given in (\ref{fxipsi})-(\ref{fconst}); and for $G_{\rm I}$, $G_{\rm V}$ and $G_{\rm VI}$  in (\ref{GI}), (\ref{GV}) and (\ref{GVI}), respectively.  Recall that $\xi$ is independent of $y$ and the common factor $f(\xi,\psi,m)$ contains (via $\psi$) all of the $y$-dependence in these weights.  This plays an important role in section \ref{factor}.

Returning to the functions $G_{\mathrm{I}}$--$G_{\mathrm{V}}$, we can also derive relations analogous to the four-point crossing matrix directly from the vertex operator formalism.  The integral of the screening charge around a small loop in the bulk containing no operators is zero. %
Deforming the contour to run along the boundary gives the blocks as described above, while picking up a complex argument of $2 \pi \alpha_- \alpha$ each time the contour passes a boundary charge $\alpha$.  Thus %
\be \label{compXRel}
0=e^{8 \pi i/\k} G_{\mathrm{I}}+e^{4 \pi i/\k} G_{\mathrm{II}}+G_{\mathrm{III}}+e^{-4 \pi i/\k} G_{\mathrm{IV}}+e^{-8 \pi i/\k} G_{\mathrm{V}}\; .
\ee
Making use of (\ref{loopeq}) then allows us to write (\ref{compXRel}) as two real equations
\bea
2 G_{\mathrm{III}}&=&(2-n^2)G_{\mathrm{I}}+nG_{\mathrm{II}}+nG_{\mathrm{IV}}+(2-n^2)G_{\mathrm{V}}\; , \quad \mathrm{and}\\
n G_{\mathrm{I}}+G_{\mathrm{IV}}&=&G_{\mathrm{II}}+nG_{\mathrm{V}}\; .
\eea
which also follow from (\ref{GI2}-\ref{GV2}).  These two equations reflect the fact that the five solutions live in a three dimensional solution space, which is consistent with the number of  configurations in the top row of figure \ref{config3}.

In the foregoing, we have simplified our figures and discussion by considering linear integration paths along the real line only.  However, our results also hold for the more general paths entwining the operators required for $\k\le4$.   Thus our solutions and the identifications of their fusion channels extend automatically to the dilute phase of the $\mathrm{O}(n)$ model, and equivalently the spin cluster behavior of the critical $Q$-state Potts models.

While the Coulomb gas formalism has considerable predictive power, our differential equation analysis has the advantage of specifying the form of the prefactor (\ref{fxipsi}), and determining the dimensionality of the entire solution space without
making a priori assumptions about the properties of the density operator.  In  addition, (\ref{DE3}) does not follow simply  in vertex operator formalism.

The three cluster configurations illustrated in figure \ref{config3} account for the three dimensional solution space for $F(\xi, \psi,m)=G(\xi,m)$ discussed above.
In addition there are three solutions that occur when the top and bottom rather than left and right sides are wired. These additional solutions are implied by the $\{\xi,\psi,m\} \leftrightarrow \{ \psi, \xi,1-m\}$ symmetry  (\ref{Sym}).  This exhausts the set of solutions that are consistent with the conformal weights and null state conditions of the correlation function  (\ref{cf1}).

%

%
%
%
\section{Factorization Behavior}  \label{factor}

In this section we consider a universal ratio that demonstrates a factorization of certain higher-order correlation functions (or the equivalent densities) in terms of less complicated correlation functions (or densities).  The ratio was originally defined at the percolation point ($\k=6$),  and is discussed in
\cite{KlebanSimmonsZiff06, SimmonsKlebanZiffPRE07, SimmonsZiffKleban08,SimmonsKleban09}.   Here we generalize it to a set of two-dimensional critical points.  The generalized ratio is universal, constant (except near the sides of the rectangle), so that factorization occurs, and independent of the vertical coordinate $y$.

In \cite{SimmonsZiffKleban08}  the universal quantity
\be \label{rho6}
\rho_{\mathrm{perc}}(x, R) =  \frac{P_{\ell r}(z,R)}{\sqrt{P_r(z,R) P_{\ell}(z,R) P_\Ho(R)}}\;, 
\ee
was considered   for critical percolation  clusters.  Here $P_\Ho(R)$ is the horizontal crossing probability given by Cardy,    and $P_{\ell r}(z,R)$ (resp.\ $P_r(z,R)$, $P_\ell(z,R)$), already introduced in subsection \ref{FKformulas}, are the densities of clusters connecting to both (resp.\ right, left) vertical sides of the rectangle.  (These densities are given in terms of $\Pi_{\AT}$,  $\Pi_{\BR}$ and $\Pi_{\BL}$  by (\ref{Pllr})-(\ref{Plrlr})  with $n = 1$.  Note that in this case $\Pi_\Ho + \Pi_\Ve = 1$.  Note also that all nonuniversal factors divide out of the ratio, so it has the same value if  cluster densities in a specific model of percolation are used in place of the CFT results.)

 The $y$-independence of $\rho_{\mathrm{perc}}$ is discussed in \cite{SimmonsZiffKleban08} and herein. 

In the limit $R \to \infty$ 
$\rho_{\mathrm{perc}}$ is constant and  $P_{\ell r}$  exactly factorizes in terms of two- and three-point functions.  We showed that in fact $\rho_{\mathrm{perc}}(\infty)=1.0299\ldots$ is the fusion coefficient for three two-leg operators.  See \cite{KlebanSimmonsZiff06, SimmonsKlebanZiffPRE07, SimmonsKleban09} for more details.

In a finite rectangle $\rho_{\mathrm{perc}}(x,R)$ is not quite constant, though it is independent of the vertical coordinate, as implied by (\ref{DE3}).  Instead $\rho_{\mathrm{perc}}(x,R)$ equals $1$ at $x=0$ and $R$, and decays exponentially, on a distance scale of the height of the rectangle, towards $\rho_{\mathrm{perc}}(\infty)$ within the bulk. Thus $\rho_{\mathrm{perc}}$ is constant to within 3\%, and the factorization  holds to good approximation for arbitrary $x$ and $R$ at the percolation point.

The results of the previous sections allow us to generalize $\rho_{\mathrm{perc}}$ to arbitrary $\k$.  To begin, we introduce the notation $\IL:=\phi_{1,2}^c(0)\phi_{1,2}^c(\mathrm{i})$, $\IR:=\phi_{1,2}^c(R)\phi_{1,2}^c(R+\mathrm{i})$, and $\sigma(z,\bar z):=\Phi_{1/2,0}(z,\bar z)$ for the wired left and right sides of the rectangle and the magnetization operator at interior point $z$ respectively.  We then generalize the definition from \cite{SimmonsZiffKleban08} as 
\be \label{rho}
\rho(z,\bar z,R)=\sqrt{%
\frac{\langle \IL{}_{[1,3]} \sigma(z,\bar z)_{[1,3]} \IR \rangle^2 \langle \IL{}_{[1,1]} \IR \rangle}%
{\langle \IL{}_{[1,1]} \sigma(z,\bar z)_{[1,3]} \IR \rangle\langle \IL{}_{[1,3]} \sigma(z,\bar z)_{[1,1]} \IR \rangle\langle \IL{}_{[1,3]} \IR \rangle}}\; ;
\ee
the motivation for this particular choice is explained below.  The subscripts  index the propagating channel between sets of operators, thereby identifying the conformal block.   The new correlations introduced here specify weights, when properly normalized these weights become densities (if $z$ dependent) or probabilities (if not).

The correlation functions $\langle \IL \,\IR \rangle$ with four $\phi_{1,2}^c$ operators, in the cases where they can be understood in terms of critical clusters, generalize  Cardy's  horizontal percolation crossing probability, as discussed in subsection \ref{pie}.  The configurations in figure \ref{crdy} determine the blocks (see (\ref{G11}-\ref{G13}) below), which are given by 
\bea \label{4opeq}
\langle \IL{}_{[\chi]} \IR \rangle &=& \langle \phi_{1,2}^c(\mathrm{i}) \phi_{1,2}^c(0){}_{[\chi]} \phi_{1,2}^c(R)\phi_{1,2}^c(\mathrm{i}+R)\rangle_{\mathcal{R}}\\ \label{4opeqb}
&=&\lim_{\e_j \to 0} \frac{\prod_{i=1}^4 \left| w'(z(u_j)) \right|^{h_{1,2}}}{(16 \e_1\e_2 \e_3 \e_4)^{h_{1,2}} } \langle \phi_{1,2}(u_4) \phi_{1,2}(u_1){}_{[\chi]} \phi_{1,2}(u_2)\phi_{1,2}(u_3) \rangle_{\mathbb{H}}\\
&=&(K')^{8 h_{1,2}}G_{\chi}(1-m)\; .
\eea
The covariance prefactor in (\ref{4opeqb}) is computed in \cite{SimmonsKleban11}.    The conformal blocks for this four point function were given in \cite{BauerBernardKytola05} for a multiple SLE process.  Due to the factors arising from the corner operators we use a slightly modified form 
\bea \label{Cardy11}
G_{1,1}(m) &=& {}_2F_1\left(2-\frac{12}{\k} ,1-\frac{4}{\k};2-\frac{8}{\k}\bigg|m\right)\\ \label{Cardy13}
G_{1,3}(m) &=&\frac{\G(12/\k-1) \G(2-8/\k)}{\G(8/\k) \G(1-4/\k)} m^{8/\k-1}{}_2F_1\left(1-\frac{4}{\k},\frac{4}{\k};\frac{8}{\k}\bigg|m\right)\; .
\eea

We   next deduce the form of the conformal blocks for the six-point function   using the methods from section \ref{Icc} and (\ref{pia})-(\ref{pib2}).  Two $\phi_{1,2}$ operators that fuse in the $\f_{1,3}$ channel cannot represent the two ends of a single boundary arc.  Also, while two $\phi_{1,2}$ operators that fuse in the identity channel do not place any conditions on which configurations contribute to the block, they do imply an extra factor of $n$ for configurations in which they are connected by a single arc.  It follows that   
\bea \label{PiAexp}
\langle \IL{}_{[1,3]} \sigma(z,\bar z)_{[1,3]} \IR \rangle&=&\Pi_\AT \phantom{+n \Pi_{\BR}}\;\,
= f(\xi,\psi,m) G_{\rm VI}(\xi,m)\\
\label{PiBrexp}
\langle \IL{}_{[1,1]} \sigma(z,\bar z)_{[1,3]} \IR \rangle&=&\Pi_\AT+n \Pi_{\BR}= f(\xi,\psi,m) G_{\mathrm{II}}(\xi,m)\\
\label{PiBlexp}
\langle \IL{}_{[1,3]} \sigma(z,\bar z)_{[1,1]} \IR \rangle&=&\Pi_\AT+n \Pi_{\BL} = f(\xi,\psi,m) G_{\mathrm{IV}}(\xi,m)\\ \label{G11}
\langle \IL{}_{[1,1]} \IR \rangle &=&\Pi_\Ho+n \Pi_\Ve\;
=(K')^{8 h_{1,2}}G_{1,1}(1-m)\\ \label{G13}
\langle \IL{}_{[1,3]} \IR \rangle  &=& \Pi_\Ho \phantom{+n \Pi_{\BR}}\;\,
=(K')^{8 h_{1,2}}G_{1,3}(1-m)\; .
\eea
One may also write
\be \label{PiVeq}
 \Pi_\Ve = (K')^{8 h_{1,2}}G_{1,3}(m) \; .
 \ee  
The method of determining these contributions via $\Pi_i$s uses $\mathrm{O}(n)$ loop model concepts, and thus applies when $8/3\le\k\le8$.  But conformal blocks are uniquely identified by the exponents in their various functional expansions, so once identified the results extend to all $\k > 0$.

The expression for $\rho(z,\bar z,R)$ thus becomes 
\bea
\rho(x, R) \label{rho2}  
&=&\sqrt{\frac{\Pi_\AT{}^2(\Pi_\Ho+n \Pi_\Ve)}{(\Pi_\AT+n \Pi_{\BR} )(\Pi_\AT+n \Pi_{\BL} )\Pi_\Ho}} \\ \label{rho3}
&=&\sqrt{\frac{ G_{\mathrm{VI}}(\xi,m)^2 \, G_{1,1}(1-m)}{G_{\mathrm{II}}(\xi,m) \, G_{\mathrm{IV}}(\xi,m) \, G_{1,3}(1-m)}} \; .
\eea
Note that the independence of $\rho(x,R)$ from the vertical coordinate $y$ is manifest.  This feature may be traced back to  (\ref{CeqfG}).  

The  particular choice  (\ref{rho}) for $\rho(x,R)$ ensures that each conformal operator  in the numerator has a counterpart in the denominator. This guarantees that we may replace each correlation function by its corresponding density, since all the nonuniversal scaling factors cancel regardless of the regularization scheme,  so that the   resulting ratio of densities is   a universal conformally invariant quantity.  We emphasize that this generalization is not unique.  It does, however, minimize the number of conformal blocks that appear in the definition, and in that sense may be considered the natural generalization of $\rho_{\mathrm{perc}}$.

The left hand sides of equations (\ref{Pllr})-(\ref{Plrlr}) and (\ref{Hxings}) then allow us to write the universal ratio $\rho$ in terms of the densities $P_i$ and the crossing probability $P_H$ for FK clusters in rectangles with independently wired vertical sides as 
\be\label{rhoFK}
\rho(x,R)=\frac{P_{\ell r}(z,R)}{\sqrt{P_\ell(z,R)P_r(z,R)P_H(R)}} \; ,
\ee
which naturally generalizes $\rho_{\text{perc}}$ in (\ref{rho6}) to $8/3 < \k < 8$.    For spin or FK clusters in rectangles with mutually wired vertical sides   one may write $\rho$ in terms in terms of the densities ${\bar P}_{\ell}(z,R),{\bar P}_r(z,R),{\bar P}_{\ell,r}(z,R)$ and crossing probability ${\bar P}_H(R)$ by using the right hand sides of  (\ref{Pllr})-(\ref{Plrlr}) and (\ref{Hxings}), respectively.
By inverting these equations, we find
\be\label{rhospin}
\rho(x,R)=\sqrt{\frac{{\bar P}_{\ell r}(z,R)^2[n^2-(n^2-1){\bar P}_H(R)]}{[(n^2-1){\bar P}_{\ell r}(z,R)-n^2{\bar P}_\ell(z,R)][(n^2-1){\bar P}_{\ell r}(z,R)-n^2{\bar P}_r(z,R)]{\bar P}_H(R)}}.
\ee
When $Q=2$, $n=1$ (dilute) and the expression (\ref{rhospin}) for $\rho$ reduces to (\ref{rhoFK}).  This occurs because the left and right sides are wired $+$ (resp.\ $-$) while the top and bottom sides are wired $-$ (resp.\ $+$) when $Q=2$.  The consequent $+/-$ duality eliminates any distinction between closing loops along the left and right sides of $\mathcal{R}$ versus closing loops across the top and bottom.   One could define $\rho$ differently so that (\ref{rhospin}) was simpler, e.g.\ by (\ref{rhoFK}), with each $P$ replaced with a ${\bar P}$.  However doing so results in an expression that does not reduce to a single OPE coefficient in the limit ${\cal R} \to \infty$ (see (\ref{rhoOPE}) below).

The ratio $\rho(x, R)$ is normalized so that $\rho = 1$ when we take $x$ to $0$ or $R$.   In terms of critical clusters, this means that as $z$ goes to the left hand side it becomes vanishingly likely that $z$ is not connected to the adjacent wired boundary and $\Pi_{\BR} \to 0$.  In addition, $\Pi_\AT \to \Pi_\Ho$ and $\Pi_{\BL} \to \Pi_\Ve$, so that indeed $\rho \to 1$ as claimed.  An analogous argument holds as $z$ goes to the right hand side.

\begin{figure}[p]
\resizebox{16cm}{4.5cm}{\includegraphics{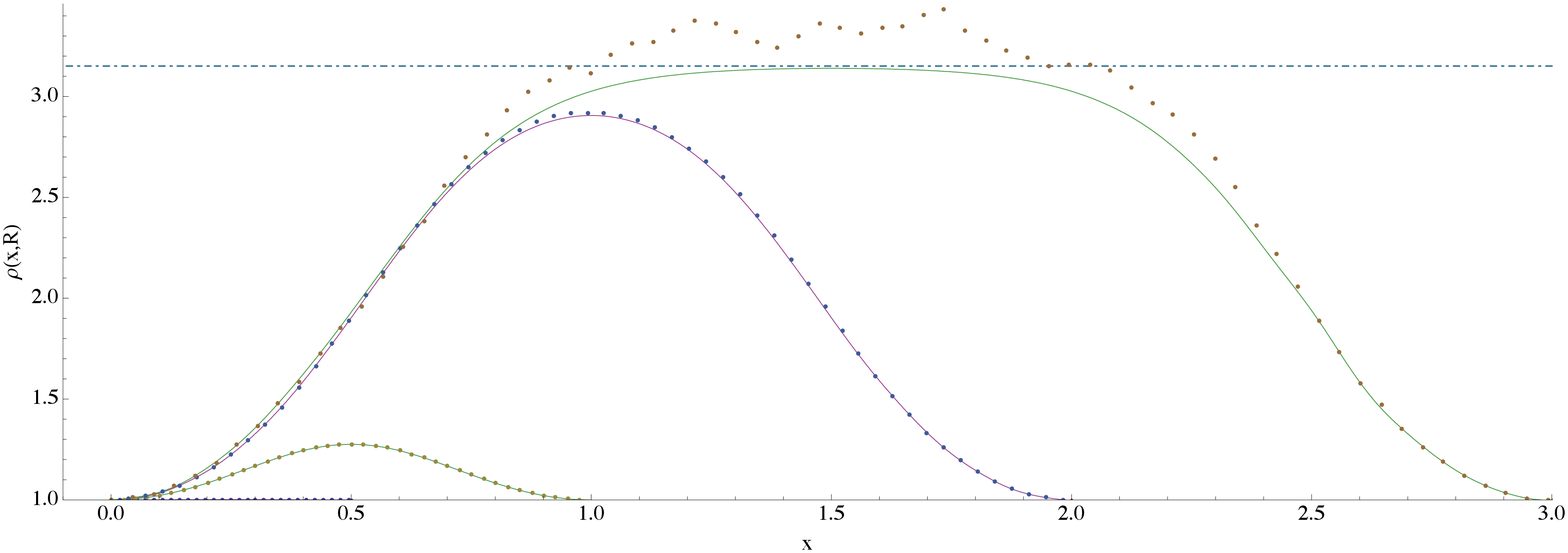}}
\caption{Numerical results for $\rho(x,R)$ for $Q=2$ spin cluster densities  ($\kappa=3$) against the CFT prediction (solid curve).  The data points are simulation values of $\rho (x,y,R)$ averaged over the $y$-coordinate with $x,R$ fixed.  Simulations were performed on a rectangle of about $10^6$ square lattice spacings.  Notice that both the theory and data plateau at $C_{1,3;1,3}^{1,3}=3.15123$ (dashed line) given in (\ref{rho_limit}).    The $R=1/2$ data are too small to show.  This data is presented separately in figure (\ref{Rhalf}).}
\label{spinQ2}
\end{figure}

\begin{figure}[p]
\resizebox{16cm}{4.5cm}{\includegraphics{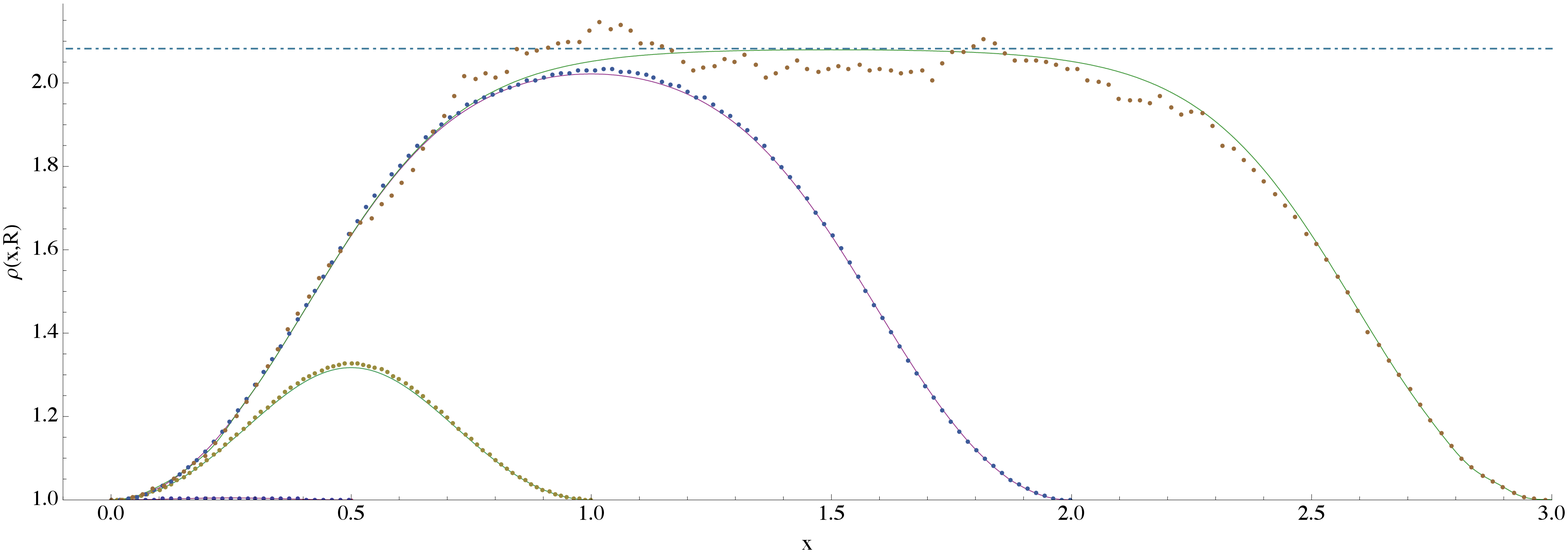}}
\caption{Same as figure \ref{spinQ2}, except for $Q=3$ spin cluster densities  ($\kappa=10/3$).  The   simulations were performed on a rectangle of about $10^6(\times4$ resp.) square lattice spacings for $R=1/2$ (resp.\ 1,\,2,\,3).  Here the plateau is at $2C_{1,3;1,3}^{1,3}/(1+\sqrt{5})=2.08229$ (dashed line) given in (\ref{rho_limit}).    See figure (\ref{Rhalf}) for expanded $R=1/2$ data.}
\label{spinQ3}
\end{figure}

We can also take the long rectangle limit $R \to \infty$ with $0 \ll x \ll R$.  Then the existence of a horizontal spanning cluster is exponentially unlikely so that $\Pi_\AT \ll \Pi_{\BR}, \Pi_{\BL}$ and $ \Pi_\Ho \ll \Pi_\Ve $.  Furthermore, we can replace the distant left and right sides with point operators, and due to the invariance with respect to the vertical direction we can assume the magnetization operator sits on the boundary.  Thus 
\bea
\rho(x, R \to \infty) &=& \frac{\Pi_\AT}{\sqrt{n \Pi_{\BR} \Pi_{\BL} \Pi_\Ho}}\\ \label{rhoOPE}
&=&\frac{\langle \phi_{1,3}(0) \phi_{1,3}(x) \phi_{1,3}(R) \rangle}{\sqrt{n \langle \phi_{1,3}(x) \phi_{1,3}(R) \rangle \langle \phi_{1,3}(0) \phi_{1,3}(x) \rangle \langle \phi_{1,3}(0)\phi_{1,3}(R) \rangle}}=\frac{C_{1,3;1,3}^{1,3}}{ \sqrt{n}}\\ \label{rho_limit}
&=&\sqrt{\hspace{.2cm}  \frac{\G\left(\frac{16-\kappa }{\kappa }\right)^2\G\left(\frac{4}{\kappa }\right)^3} {n(\k) \, \G\left(\frac{8-\kappa }{\kappa }\right)\G\left(\frac{12-\kappa }{\kappa }\right)\G\left(\frac{8}{\kappa }\right)^3}}
\eea
where the final line uses the known expression for the OPE coefficient as given in \cite{SimmonsKleban07}.  (Recall that $n(\kappa)$ is given in (\ref{loopeq}).)  Note that the value of $C_{1,3;1,3}^{1,3}$ increases monotonically from $0$ to $5 \sqrt{2}$ as $\k$ decreases from $8$ to $2$. Some specific values are given in the captions to figures \ref{fkQ2}-\ref{spinQ3}.  For percolation ($\k = 6$) $C_{1,3;1,3}^{1,3}$ is close to $1$, and the factorization is similar anywhere in the rectangle, since $\rho = 1$ at the edges.  When $C_{1,3;1,3}^{1,3}$ is far from $1$, however, the nature of the factorization is different near the edges and far from them.

Note that the long rectangle limit, as described above, fails when $\k=8/3$.  This value of $\k$ corresponds to the dilute phase with $n=0$ and from (\ref{rho2}) it follows that $\rho(x,R)=1$, in apparent contradiction to (\ref{rho_limit}). This is because the two limits $\k \to 8/3$ and $R \to \infty$ do not commute; vertical crossings are suppressed as $n \to 0$, while horizontal crossings are suppressed as $R \to \infty$.   This problem does not occur in the dense phase with $n=0$ ($\k=8$).  Here the limits do commute and $\rho=1$, because the curves are space filling, so there is no suppression of horizontal crossings for large $R$.

\begin{figure}[p]
\resizebox{16cm}{4.5cm}{\includegraphics{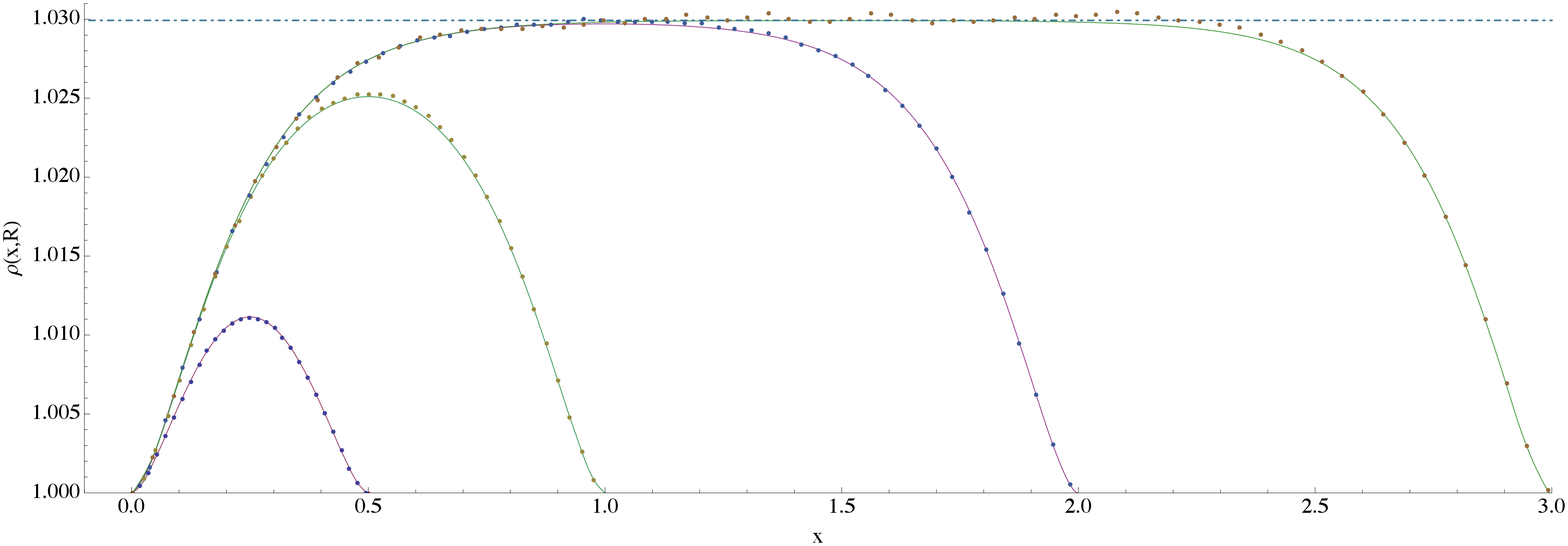}}
\resizebox{16cm}{4.5cm}{\includegraphics{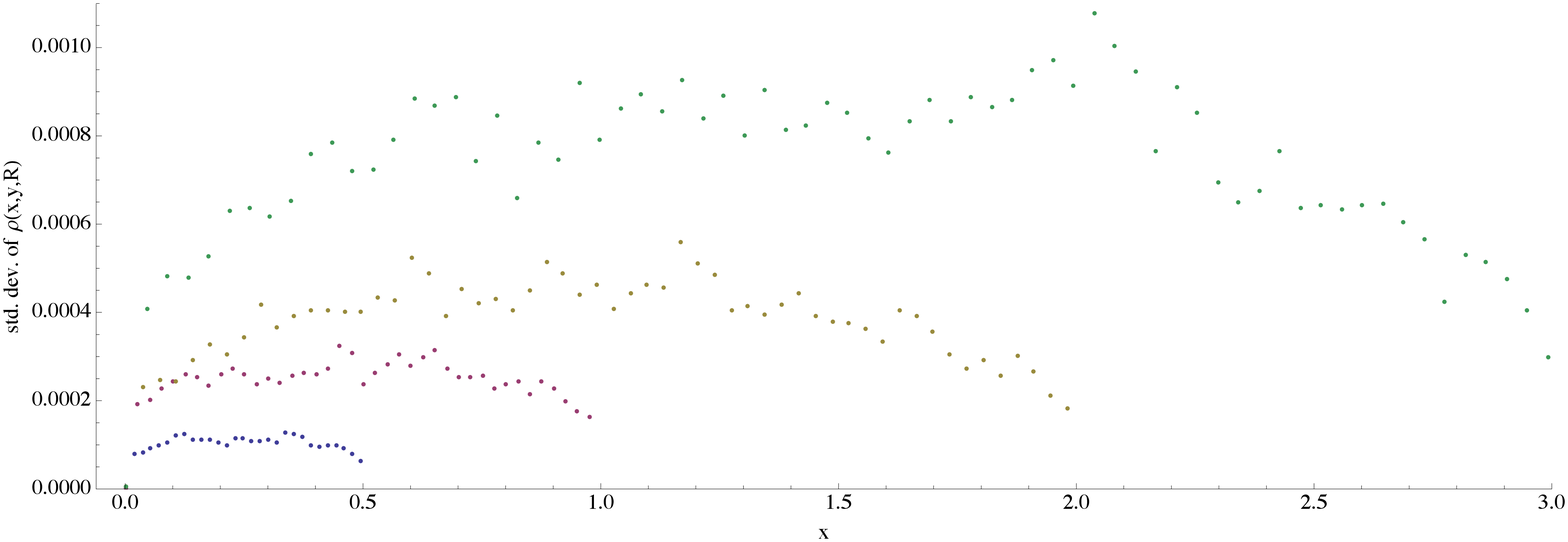}}
\caption{Upper graph:  Numerical results for $\rho(x,R)$ for percolation ($\kappa=6$) against the CFT prediction (solid curve).  The data points are simulation values of $\rho (x,y,R)$ averaged over the $y$-coordinate with $x,R$ fixed.    Simulations were performed on a rectangle of about $10^6$ square lattice spacings.  Notice that both the theory and data plateau at $C_{1,3;1,3}^{1,3}=1.02993$ (dashed line) given in (\ref{rho_limit}).  The lower graph shows, at fixed $x,R$, the standard deviation of measured $\rho(x,y,R)$ values  from their average along the height of the rectangle.}
\label{Q1}
\end{figure}

\begin{figure}[p]
\resizebox{16cm}{4.5cm}{\includegraphics{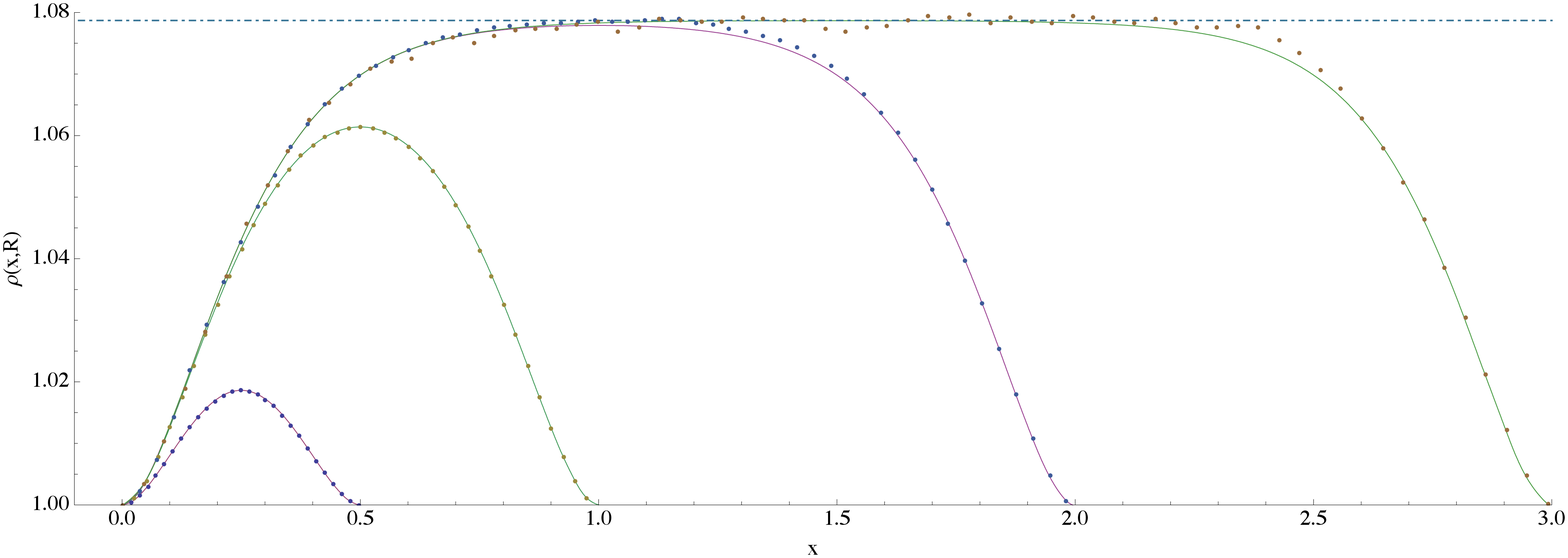}}
\caption{Same as upper figure \ref{Q1}, except for $Q=2$ FK cluster densities ($\kappa=16/3$).  Here the plateau is at $C_{1,3;1,3}^{1,3}/2^{1/4}=1.07871$ (dashed line), from (\ref{rho_limit}).}
\label{fkQ2}
\end{figure}

\begin{figure}[p]
\resizebox{16cm}{4.5cm}{\includegraphics{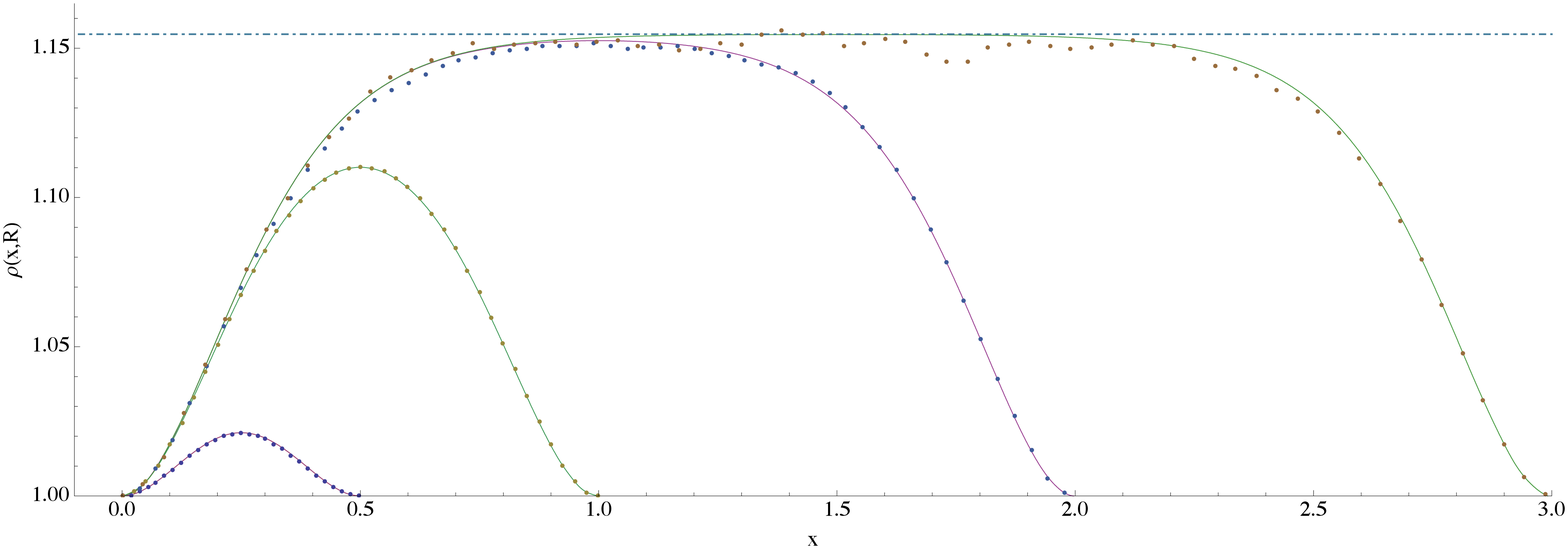}}
\caption{Same as upper figure \ref{Q1}, except for $Q=3$ FK cluster densities ($\kappa=24/5$).  The   simulations were performed on a rectangle of about $10^6(\times4$ resp.) square lattice spacings for $R=1/2$ (resp.\ 1,\,2,\,3).  Here the plateau is at $C_{1,3;1,3}^{1,3}/3^{1/4}=1.15470$ (dashed line), from (\ref{rho_limit}).}
\label{fkQ3}
\end{figure}

The form of this ratio for the $Q=1,2$ and $3$ state Potts models   (for either FK or spin clusters) is given in figures \ref{Q1}, \ref{fkQ2}, \ref{fkQ3}, \ref{spinQ2} and \ref{spinQ3} (solid curves) for various values of the aspect ratio $R$.
Note   that for $R = 2$ and $3$ there is virtually no difference between the $\rho(x,R)$ curves for a given $Q$ value when $0<x<1$,   although the spin clusters deviate slightly more.    This is consistent with the observation made in \cite{SimmonsZiffKleban08}   for percolation ($Q=1$)   that the semi-infinite strip result $\rho(x,\infty)$ is a good approximation to $\rho(x,R)$ for $0<x<R/2$ and $R\gg1$.

Note that taking the limit $R \to \infty$ with $0 \ll x \ll R$, in which case $\rho(x,R)$ becomes a constant, is equivalent to bringing the pair of $\phi_{1,2}$ operators in (\ref{cf1}) on the left hand side of the rectangle together, and likewise the pair on the right.  Applying this to $\langle \IL{}_{[1,3]} \sigma(z,\bar z)_{[1,3]} \IR \rangle$ one obtains a four-point correlation function with two $\phi_{1,3}$ operators; the other correlation functions in (\ref{rho}) can be treated similarly.  The result  generalizes our exact percolation factorization formula for the densities of clusters anchored to two points \cite{SimmonsKlebanZiffPRE07} to all the critical models mentioned here.  By keeping $x$ near $0$, one could similarly obtain results for the densities of clusters anchored to one interval and one point.  We will examine the consequences of these observations elsewhere.


\section{Density of percolation crossing clusters} \label{Dpcc}

We now restrict our attention to $\k=6$, or critical percolation.  We return to percolation, not just for its intrinsic interest, but because we are now in a position to exploit the property known as locality in the SLE literature \cite{LawlerSchrammWerner01}. This property allows us to use the results from the subsection \ref{Icc}, with wired boundary conditions on the vertical sides, to   derive an explicit expression for the density of clusters that cross a rectangle with free boundary conditions.

With respect to a growing SLE hull, `locality' means that as the curve grows in time, its statistics  are completely independent of boundaries up until the hull actually touches one.  Percolation exhibits this property because the decision to include or exclude a link is determined by a purely local random variable and not by the state  of neighboring sites.  This implies that the bulk  sites all  contribute a fixed amount to the overall weight of a configuration regardless of the state of the boundary sites.

It is this insensitivity to the boundary conditions that allows us to generalize the idea of wiring.  Suppose we condition boundary sites  in some way, and find the density of clusters attached to certain segments, as above.  If we change the conditioning and measure the density again, then the two densities are directly comparable due to the fixed contribution of the bulk sites to the overall weight.

This is intrinsically different from a system without locality. In general conditioning a set of boundary sites changes the degrees of freedom of the attached clusters, so boundary conditions are a defining characteristic of the system.  For a system with locality we needn't think of boundary conditions as fixed parameters; instead they provide a variable means of probing the connectivities of clusters within the system. 

\begin{figure}[htbp] 
   \centering
   \includegraphics[width=.56 \columnwidth]{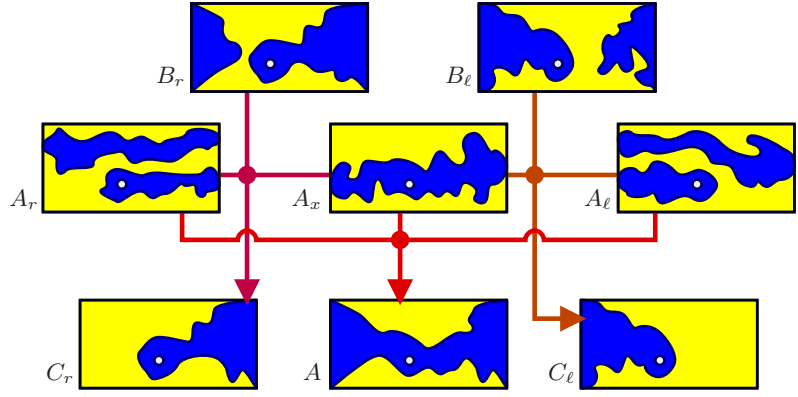}
   \hspace{-.6\columnwidth}
      \makebox[.03 \columnwidth]{\raisebox{18ex}{$\AR$}}
   \hspace{-2ex}
      \makebox[.03 \columnwidth]{\raisebox{1ex}{$\CR$}}
   \hspace{5.25ex}
      \makebox[.03 \columnwidth]{\raisebox{30ex}{$\BR$}}
   \hspace{7.75ex}
      \makebox[.03 \columnwidth]{\raisebox{18ex}{$\AX$}}
   \hspace{-5ex}
      \makebox[.03 \columnwidth]{\raisebox{1ex}{$\AT$}}
   \hspace{8.75ex}
      \makebox[.03 \columnwidth]{\raisebox{30ex}{$\BL$}}
   \hspace{4.5ex}
      \makebox[.03 \columnwidth]{\raisebox{1ex}{$\CL$}} 
   \hspace{-2ex}
      \makebox[.03 \columnwidth]{\raisebox{18ex}{$\AL$}}
   \hspace{20ex}
   \caption{Percolation configurations discussed in this section.  We use fixed boundary conditions on both (resp.\ right, left) vertical sides to calculate the densities corresponding to $\BR$, $\BL$, $\AT$ (resp.\ $\CR$,  $\CL$).   Due to locality, however, this decomposition of the densities holds regardless of boundary conditions.   The two top rows represent the disjoint subsets that span the configuration space.  The bottom row represents quantities that we calculate here which are composites, as indicated by the paths from their constituents in the upper rows. }
   \label{config}
\end{figure}

We begin with the three   configurations   illustrated in figure \ref{config3}, and proceed to decompose the contributions to $\Pi_\AT$, as in figure \ref{config}.
We distinguish configurations according to whether the point $z$ actually belongs to a crossing cluster, as in configuration type $\AX$, or whether it connects to the crossing cluster only through the wired boundaries on the right or left hand side, as in type $\AR$ and $\AL$ respectively.    Our goal is to determine the density for configurations of type $\AX$, since these actually connect both vertical sides of the rectangle, even when they are not wired.  They also constitute the horizontal crossing clusters for a rectangle with free boundary conditions on all sides. Thus
\be \label{PAeq}
 P_A = P_{\AR}+P_{\AL}+P_{\AX}\; .
\ee

For percolation, the expressions for the conformal blocks $G$ in (\ref{GI})-(\ref{GVI}) are equivalent to those  calculated in \cite{SimmonsZiffKleban08}, however   (\ref{fxipsi})  gives us an expression for the prefactor $f(\xi,\psi,m)$,   which contains the explicit vertical dependence not previously available, thus allowing a complete calculation of the densities.

Now we consider the correlation functions given by wiring either the left or right hand sides of the rectangle.  In \cite{KlebanSimmonsZiff06} we calculated the density of   percolation   clusters at a point $w=u+\ci v$ in the upper half-plane that are attached to an interval on the real line $I=(u_1,u_2)$,
\be
P_I(w)   \sim |w-\bar w|^{-5/48}\left(2\, \Imag\left[\eta^{1/4}\right]\right)^{1/3}\; ; \quad \eta=\frac{(\bar w-u_1)(w -u_2)}{(w-u_1)(\bar w -u_2)}  \; .
\ee
  Note that for percolation the normalization factor $\Pi_H+ n \Pi_V = 1$, so the weights found in \cite{KlebanSimmonsZiff06} and also those from (\ref{cf1})  are in fact densities.

We   transform $P_I(w)$ into    the rectangle ${\cal R}$ using the mapping  $w(z)=m\,\je{sn}{z}^2$ from \cite{SimmonsKleban11}.    The interval $\{-\infty,0\}$ will map to the left side and $\{m,1\}$ maps to the right side, thus the corresponding cross-ratios are
\bea
\eta_L &=& \frac{w}{\bar w}=\exp\left(4 \ci\, \arg(\je{sn}{z}) \right)\\
\eta_R &=& \frac{(\bar w-m)(w -1)}{(w-m)(\bar w -1)} = \exp\left(4 \ci\, \arg(\je{dc}{z})\right)\; .
\eea
This   gives, if the right or left side of the rectangle is wired, respectively
\bea
P_\CR(z,\bar z) &=& 2^{1/3}(K')^{5/48}\sin^{1/3}\left[ \arg(\je{dc}{z}) \right]\left(\frac{\Imag\left[\je{sn}{z}^2\right]}{\left| \je{sn}{z} \je{cn}{z} \je{dn}{z} \right|} \right)^{-5/48}\\
P_\CL(z,\bar z) &=& 2^{1/3}(K')^{5/48}\sin^{1/3}\left[\arg(\je{sn}{z}) \right]\left(\frac{\Imag\left[\je{sn}{z}^2\right]}{\left| \je{sn}{z} \je{cn}{z} \je{dn}{z} \right|} \right)^{-5/48}\; .
\eea    
The coefficient $2^{1/3}= 2^{ h_{1,3}}$ indicates the presence of a corner operator, as in (\ref{fconst}).

Using identities for elliptic functions   and (\ref{fxipsi})   we can rewrite these quantities in terms of   the real coordinates (\ref{xipsidef}): 
\bea \nonumber
P_\CR(z,\bar z) &=&f(\xi,\psi,m)
\left[\mathrm{dc}^2(x K'|m)+m\,\mathrm{cd}^2(x K'|m)-\mathrm{dn}^2(y K'|1-m)-m\, \mathrm{nd}^2(y K'|1-m)\right]^{1/6}\\ \nonumber
&=&  2^{1/3} (K')^{5/48}\left[\frac{(1-m \xi^2)^2}{ \xi (1-\xi) (1-m \xi)} + \frac{(1-(1-m) \psi^2)^2}{ \psi (1-\psi) (1-(1-m) \psi)} -4 \right]^{-11/96}     \\ \nonumber
&\times& \left[ \frac{1-m\, \xi}{1-\xi}+  m\frac{1- \xi}{1-m\, \xi}    -\frac{(1-(1-m)\y)^2+m}{1-(1-m)\y}\right]^{1/6}  \\ \nonumber 
P_\CL(z,\bar z) 
&=&f(\xi,\psi,m)
\left[\mathrm{ns}^2(x K'|m)+m\,\mathrm{sn}^2(x K'|m)-\mathrm{dn}^2(y K'|1-m)-m\, \mathrm{nd}^2(y K'|1-m)\right]^{1/6}\\ \nonumber
&=&   2^{1/3} (K')^{5/48}\left[\frac{(1-m \xi^2)^2}{ \xi (1-\xi) (1-m \xi)} + \frac{(1-(1-m) \psi^2)^2}{ \psi (1-\psi) (1-(1-m) \psi)} -4 \right]^{-11/96}   
\left[\frac{1}{\xi}+m\, \xi-\frac{(1-(1-m)\y)^2+m}{1-(1-m)\y}\right]^{1/6}\; .
\eea
  Note that on use of (\ref{Sym}) and the invariances of $f(\xi,\psi,m)$ \cite{SimmonsKleban11}, the (final) expressions for $P_\CR$ and $P_\CL$ transform  properly under mirror symmetry about $x=R/2$ or $y=1/2$.

With these expressions for the densities $P_\CR$ and $P_\CL$ we can isolate the density of crossing clusters $P_{\AX}$.               Now  
\bea \label{PCrexp}
P_\CR&=&P_{\BR}+P_{\AR}+P_{\AX} \;, \quad \mathrm{and} \\
P_\CL&=&P_{\BL}+P_{\AL}+P_{\AX} \;.  \label{PClexp}
\eea  
Note that in (\ref{PCrexp}) and (\ref{PClexp})   the left and and right hand sides involve quantities originally defined with different boundary conditions.  Only when there is locality does this difference become irrelevant.

We now check that the normalizations of the various quantities are consistent.  Using (\ref{PiBlexp}), (\ref{PClexp}) and figure \ref{config}, we find $fG_{\mathrm{IV}} - P_\CL =  P_{A_r}$.  The latter is the density of points $z$  that connect to the right hand side but not to the left, in configurations with a crossing cluster.  As  $x \to 0$, our expression for this vanishes, as it must.  Using  (\ref{PiBrexp}) and (\ref{PCrexp}) we come to the same conclusion for $f \, G_{\mathrm{II}} - P_\CR = P_{A_l}$ as $x \to R$.   

 Thus  we arrive at an explicit formula for the density of clusters that cross the rectangle horizontally regardless of wiring at the sides  (equivalently, cross a rectangle horizontally with open boundaries).
\begin{align} \label{PAxexp}
P_{\AX}&=P_\CR+P_\CL-P_\AT-P_{\BR}-P_{\BL} \\ \nonumber
&=   2^{1/3} (K')^{5/48}\left[\frac{(1-m \xi^2)^2}{ \xi (1-\xi) (1-m \xi)} + \frac{(1-(1-m) \psi^2)^2}{ \psi (1-\psi) (1-(1-m) \psi)} -4 \right]^{-11/96}     \\ \label{xingdens}
&\hspace{1in}
\Bigg[ 
\left(\frac{1-m\, \xi}{1-\xi}+   m\frac{1- \xi}{1-m\, \xi}    -\frac{(1-(1-m)\y)^2+m}{1-(1-m)\y}\right)^{1/6}+\left(\frac{1}{\xi}+m\, \xi-\frac{(1-(1-m)\y)^2+m}{1-(1-m)\y}\right)^{1/6}\\ \nonumber
&\hspace{1in}
-\frac{(1-m)^{1/3 }}{\left[\xi (1-\xi )(1-m \xi )\right]^{1/6}}F_1\left(\frac{1}{3};\frac{2}{3},-\frac{2}{3};\frac{2}{3}\bigg| m,m \xi \right) \Bigg]\; . 
\end{align}
  (The invariance of the last term in the bracket in (\ref{xingdens}) under mirror symmetry about $x=R/2$ follows from (\ref{Sym}) and standard results for the Appell functions.)  \\

\begin{figure}[p]
\begin{tabular}{|c|c|c|c|c|}
\hline 
& \hspace{.5cm}$R=1/2$\hspace{.5cm} & \hspace{.5cm}$R=1$\hspace{.5cm} & \hspace{.5cm}$R=2$\hspace{.5cm} & \hspace{.5cm}$R=3$\hspace{.5cm} \\
\hline
$Q=1$ & 0.00010 & 0.00024 & 0.00038 & 0.00074\\
\hline
FK $Q=2$ & 0.00013 & 0.00043 & 0.00079 & 0.0025\\
\hline
FK $Q=3$ & 0.00018 & 0.00061 & 0.0022 & 0.0063\\
\hline
spin $Q=2$ & 0.000058 & 0.0013 & 0.0096 & 0.16\\
\hline
spin $Q=3$ & 0.00012 & 0.0010 & 0.023 & 0.092\\
\hline
\end{tabular}
\caption{Standard deviation of measured $\rho$ values for figures \ref{Q1} and \ref{fkQ2}-\ref{Rhalf}  at fixed $x$ and $R$ from the measured averages, as in the lower graph of figure \ref{Q1} (percolation).    The average of these standard deviations over $x$ is shown. In each case, the graph of standard deviation vs.\ $x$ resembles the the lower graph in figure \ref{Q1}, but the magnitude varies.}
\label{stdevtable}
\end{figure}


\begin{figure}[p]
\resizebox{8.5cm}{5cm}{\includegraphics{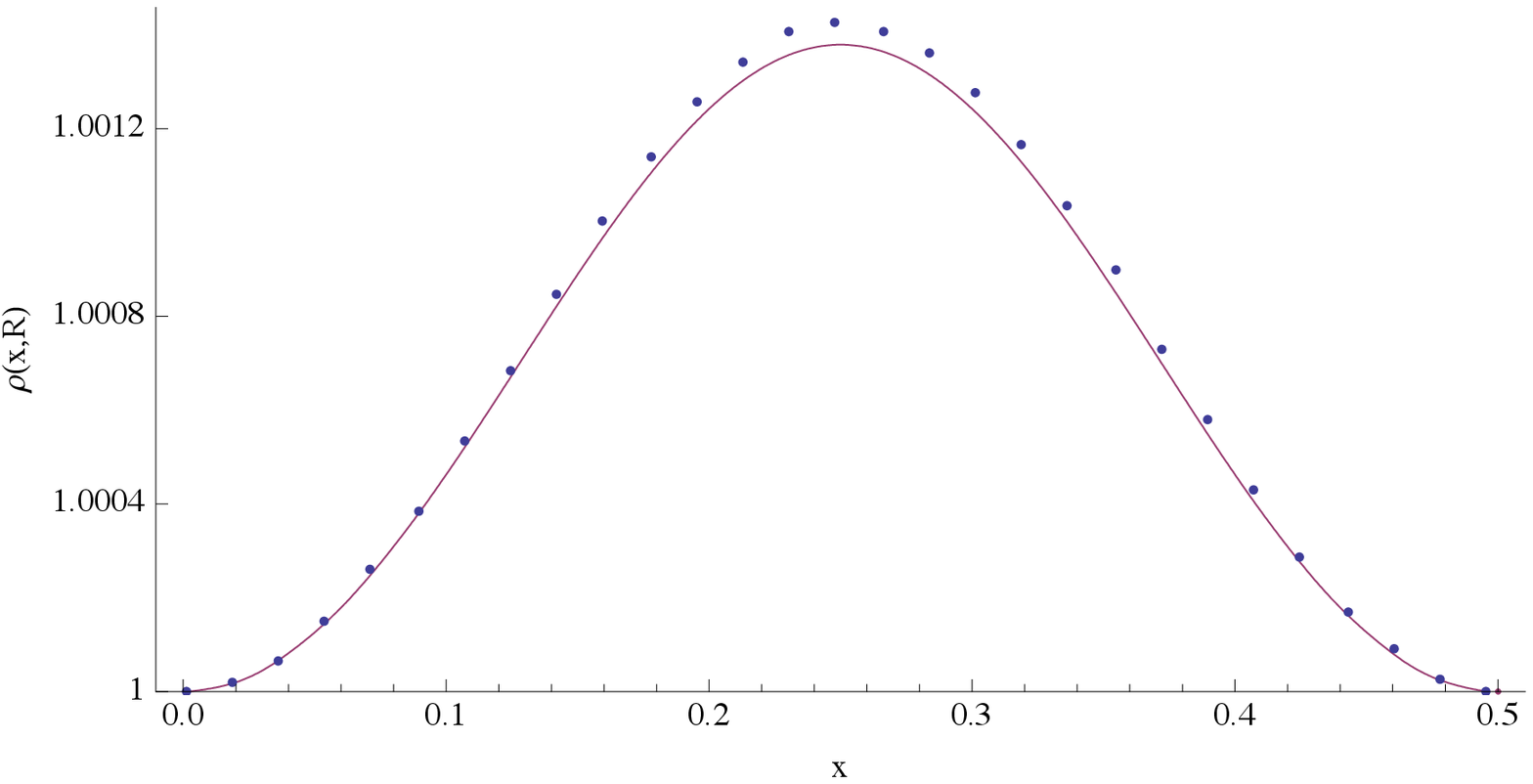}}
\resizebox{8.5cm}{5cm}{\includegraphics{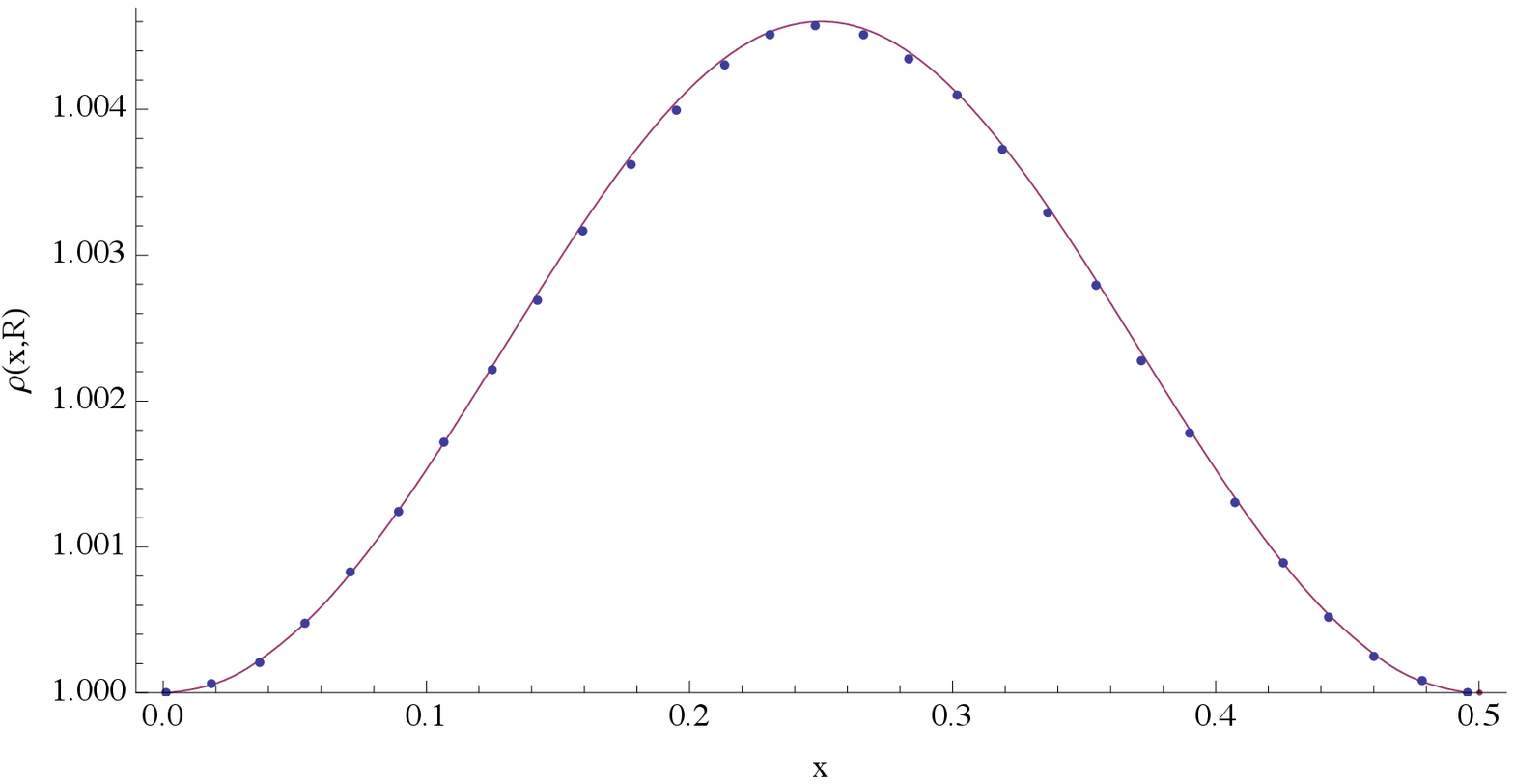}}
\caption{Results for $\rho(x,R=1/2)$ for $Q=2$ (left) and $Q=3$ (right) spin cluster densities.  The CFT prediction is the solid curve.  The $Q=2$ (resp.\ 3) simulation was performed on a rectangle with area about equal to $10^6(\times4$ resp.) square lattice spacings.}
\label{Rhalf}
\end{figure}

\begin{figure}[p]
\resizebox{12cm}{6cm}{\includegraphics{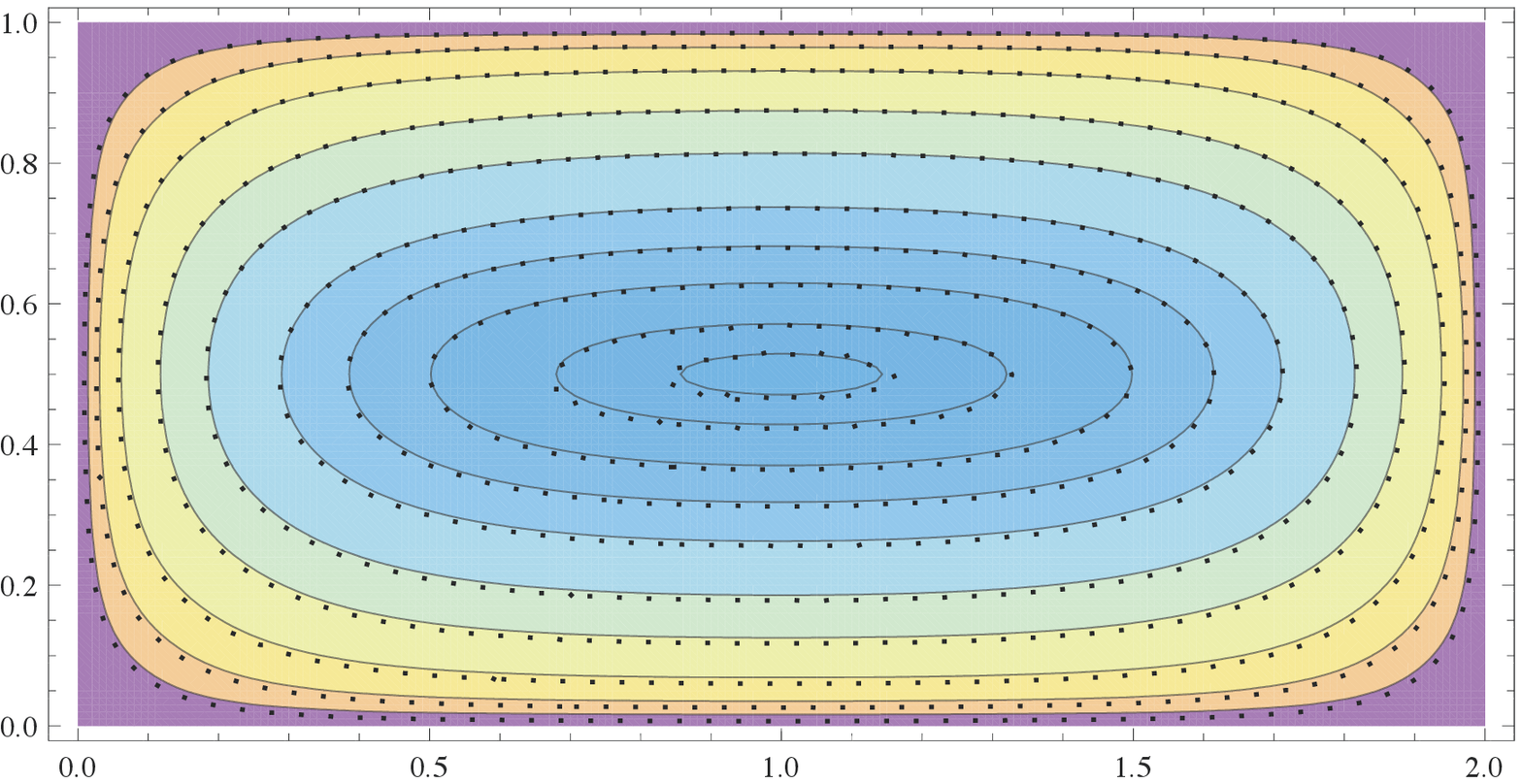}}
\caption{Density of percolation crossing clusters  at $R=2$.  Solid curves are predictions from (\ref{xingdens}), data points from simulations. Data deviates more from the prediction  near the center  since $P_{A_x}$ is almost flat there.  See text in section \ref{Dfbc} for values.}
\label{PiAxplots}
\end{figure}


\section{Simulation results for $\rho(x,R)$ and $P_{A_x}$ \label{Sims}}

In this section, we present numerical results that verify the predictions (\ref{pia}-\ref{pib2}) via calculation of the universal ratio $\rho(x,R)$ (see (\ref{rho2})).  Here we simulated the $Q=1,\, 2,$ and  $3$-state Potts models on a square lattice in rectangles $\mathcal{R}$ of aspect ratios $R=1/2,\, 1,\, 2,$ and  $3$ using the Swendsen Wang (SW) algorithm \cite{sw} at the critical bond activation probability $p_c(Q)=\sqrt{Q}/(1+\sqrt{Q})$.  Both FK and spin clusters were considered, and we found good agreement between the predictions of this paper and simulation.  We also verify (\ref{xingdens}), the density $P_{A_x}$ of percolation clusters that cross horizontally with free boundary conditions via simulations.  \\

\subsection{$Q=1$: Percolation}

\subsubsection{The universal ratio $\rho(x,R)$}

When $Q=1$, the rectangle $\mathcal{R}$ is filled by a single spin cluster in one spin state.  The FK clusters within it are bond percolation clusters, and because $p_c(1)=1/2$, the SW algorithm is equivalent to simulating critical bond percolation.  

The boundary conditions are either ``fixed" (``wired") or free (``open").   All bonds on a ``fixed" interval are activated; on a ``free" interval they are activated with probability $1/2$ (just as in the interior of $\mathcal{R}$).   The insertion of $\phi_{1,2}$  changes the boundary condition from fixed to free, so the boundary conditions on the sides of $\mathcal{R}$ alternate between fixed and free. We choose to wire the left and right sides, so the top and bottom are free.

In our simulations, we measured the density $P_\ell(z,R)$ (resp.\ $P_r(z,R),P_{\ell r}(z,R)$) of clusters touching $z\in\mathcal{R}$ that are anchored to the left (resp.\ right, left and right) side(s) of $\mathcal{R}$.  The density was found numerically by dividing the number of samples where $z$ is connected to the left (resp.\ right, left and right) side(s)  through activated bonds by the number of samples.  We also measured the probability of horizontal cluster crossings $P_H(R)$.  The length and width of $\mathcal{R}$ were chosen so that the area of its interior is approximately equal to $10^6$ square lattice spacings.  When $R=1/2$ or $1$, $5\times10^6$ samples were generated, and when $R=2$ or $3$, $15\times10^6$ samples were generated.

In figure \ref{Q1}, we plot simulation results for $\rho_{\text{perc}}(x,R)$ and the CFT prediction in (\ref{rho6}).  We expect $\rho_{\text{perc}}$ to be independent of its $y$-coordinate, so in the first plot, we average the measured  $\rho_{\text{perc}}$ values over the $y$-coordinate for fixed $x$ and plot the average as a function of the $x$-coordinate, as well as its standard deviation from the measured values. When $x$ nears 0 or $R$, $\rho$ decays exponentially to 1.

Some of our percolation results overlap with the simulations in \cite{SimmonsZiffKleban08}.  However, we include them here for ease of comparison with our results for Potts models at $Q = 2$ and $3$, which are new.

\subsection{$Q=2$ and $3$-state Potts Models}

Here, two different types of clusters may be considered: FK clusters and spin clusters.  In the continuum limit, the boundaries of FK clusters anchored to the sides of $\mathcal{R}$ are SLE curves in the dense phase with speed $\kappa$ related to $Q = n^2$ via (\ref{loopeq}).  When $Q=2,\kappa=16/3$, and when $Q=3$, $\kappa=24/5$.  On the other hand, the boundaries of spin clusters are SLE curves in the dilute phase with speed $\kappa'$.  As described in subsection \ref{pie}, $\kappa'=16/\kappa$, where $\kappa$ is the speed of the SLE curve for the corresponding FK cluster boundaries.  When $Q=2,\kappa'=3$, and when $Q=3$, $\kappa'=10/3$.

The boundary conditions to be used will depend on which type of cluster, FK or spin, we wish to study.  This is because the boundary condition change induced by the insertion of $\phi_{1,2}$ operators at the corners of the rectangles is different in the dense phase from that in the dilute phase.  Consequently there are two different ways to implement the boundary conditions, as discussed below.

\subsubsection{FK clusters: the universal ratio $\rho(x,R)$}

In simulations of the $Q$-state Potts model in a rectangle $\mathcal{R}$ with independently wired left/right sides, we measured the density $P_{\ell}(z,R)$ (resp.\ $P_r(z,R),P_{\ell,r}(z,R)$) of FK clusters at lattice site $z=x+{\rm i}\, y\in\mathcal{R}$ connected to the left (resp.\ right, left and right) side(s)  of $\mathcal{R}$.  These densities were found by dividing the number of samples where the lattice site $z$ touches an FK cluster anchored to the left (resp.\ right, left and right) sides by the number of samples.   In the same simulations, we also measured the probability $P_H(R)$ of a horizontal FK cluster crossing (which can occur only in samples where the left and right sides have the same spin), given by (\ref{Hxings}).  The densities are given in (\ref{Pllr})-(\ref{Plrlr}), and the  probability $P_H(R)$  in (\ref{rhoFK}). 

As discussed in subsection \ref{pie}, $\phi_{1,2}$ changes the boundary condition at the corners of $\mathcal{R}$ from fixed to free in the dense phase.  In the FK representation, ``fixed" implies that all bonds along the left and right sides of $\mathcal{R}$ are activated in each sample, and ``free" implies that bonds along the top and bottom sides are activated with probability $p_c(Q)$.  To ensure that these boundary conditions are respected, we modify the Swendsen-Wang algorithm \cite{sw} by laying a bond with probability 1 between all adjacent sites of the left and right sides before each update.

The system size and sample size were chosen differently for $Q=2$ and $Q=3$.   When $Q=2$, the length and width of $\mathcal{R}$ were chosen so that the area of its interior is approximately equal to $10^6$ square lattice spacings.  $5\times10^6$ samples were generated when $R=1/2$ or $1$, and $15\times10^6$ samples were generated when $R=2$ or $3$.   When $Q=3$, we noticed that the results for the universal ratio $\rho$ (\ref{rhoFK})  in a rectangle of $10^6$ square lattice spacings were noticeably lower than the theory prediction, so in this case we quadrupled the area of $\mathcal{R}$ to about $4\times10^6$ square lattice spacings  when $R=1$, $2$ or $3$, which improved the agreement.  For $Q=3$, $5\times10^6$ samples were generated when $R=1/2$, and $15\times10^6$ samples were generated when $R=1,\,2$ or $3$.

In figure \ref{fkQ2}, we plot values of $\rho$ measured in our simulations for $Q=2$ ($\kappa=16/3$) and observe good agreement.  We expect $\rho$ to be independent of its $y$-coordinate, so we average $\rho$ over the $y$-coordinate for fixed $x$ and plot the average as a function of the $x$-coordinate.    In figure \ref{fkQ2} (resp.\ \ref{fkQ3}), we plot values of $\rho$ measured in our $Q=2$ ($\kappa =16/3$) (resp. $Q=3$ ($\kappa=24/5$)) simulations.  Also for fixed $x$, we measured the standard deviation of the measured $\rho(x,y)$ from the corresponding measured average over $y$.  The average of these standard deviations over all $x$ values is shown in the table in figure \ref{stdevtable}.  For small $R$, these deviations are less than $0.013\%$, and for larger $R$, they are less than $0.63\%$.   Notice that when $R$ is large ($\approx3$), $\rho$ is roughly constant except near the sides of the rectangle  and equal to the fusion coefficient divided by $\sqrt{n}$ (\ref{rho_limit}).  When $x$ nears 0 or $R$, $\rho$ decays exponentially to 1.

\subsubsection{Spin Clusters: the universal ratio $\rho(x,R)$}

In simulations of the $Q$-state Potts model in $\mathcal{R}$ with mutually fixed left/right sides, we measured the density  ${\bar P}_{\ell}(z,R)$ (resp.\ ${\bar P}_r(z,R),{\bar P}_{\ell,r}(z,R)$)  of spin clusters at lattice sites $z=x+{\rm i}\, y\in\mathcal{R}$ connected to the left (resp.\ right, left and right) side(s) of $\mathcal{R}$.  These densities were found by dividing the number of samples where the lattice site $z$ is in a spin cluster anchored to the left (resp.\ right, left and right) side(s)  by the number of samples.  We also measured the probability of horizontal spin cluster crossings ${\bar P}_H(R)$.

As mentioned in subsection \ref{pie}, $\phi_{1,2}$ in the dilute phase changes the boundary condition at the corners of $\mathcal{R}$ from spin type A (left and right sides) to an unbiased mix of spins that are not A (top and bottom sides).  Notice that both the left and right sides are necessarily mutually wired, in contrast with FK clusters.  

Because we cannot allow a spin cluster of type A to touch the top and bottom of the rectangle, how to maintain the boundary condition change at the corners with each update is different from that of FK clusters.  Suppose that we start with some arbitrary spin configuration inside $\mathcal{R}$ that satisfies the necessary boundary conditions.  To update the system, we activate all bonds on the left side and grow the FK cluster anchored to it via the SW algorithm.  Note that this FK cluster cannot touch the top and bottom since these sides do not contain type A spins.  If the FK cluster strikes the right side, the whole right side is absorbed into it because the right side is wired.  If it doesn't, then we repeat this process on the right side, and we update the spins in the FK clusters anchored to either side to type A.  Next, we grow the FK clusters anchored to the top and bottom, and their spins are updated to a type other than A with uniform probability.  Finally, we grow an FK cluster from any site in the interior of $\mathcal{R}$ whose spin is not yet updated, with no restriction on the updated spin type.

For given $Q$ and $R$, the rectangle size and number of samples were chosen in the same way as for $Q=2, \,3$ FK clusters.  When $Q=3$, we noted that the results for  the universal ratio $\rho$ (\ref{rhospin})  in a rectangle of $10^6$ lattice spacings were noticeably greater than the prediction, in contrast to FK clusters.  So to obtain better results for $Q=3$, we again quadrupled the area of $\mathcal{R}$  when $R=1$, $2$ or $3$.

In figure \ref{spinQ2} (resp.\ \ref{spinQ3}), we plot values of $\rho$ measured in our $Q=2$ ($\kappa =3$) (resp. $Q=3$ ($\kappa=10/3$)) simulations.  Notice that the  data  at $R=1/2$ is too small to see.    It is plotted on a more visible scale in (\ref{Rhalf}).  As done earlier, we plot averages of the measured $\rho$ over the $y$-coordinate for fixed $x$.  Standard deviations of the measured $\rho(x,y)$ from the corresponding measured average over $y$ are shown in the table in table \ref{stdevtable}.  For small $R$, these deviations are less than $0.011\%$, and for larger $R$, they are less than $15\%$. We comment on possible reasons for large deviations in the latter case below.   Again, when $R$ is large $(\approx3)$, $\rho$ is roughly constant and equal to the fusion coefficient divided by $\sqrt{n}$ (\ref{rho_limit}).  When $x$ nears 0 or $R$, $\rho$ decays exponentially to 1.  

The standard deviations for spin clusters at large $R$ shown in table \ref{stdevtable} are noticeably greater than their FK counterparts.  Indeed,  figures \ref{spinQ2} and \ref{spinQ3} show deviations from the mean value of $\rho(R=3)$ greater than those in the corresponding FK cases.  This may be attributed to the increase of conformal weight $h_{1,3}$.  In the large $R$ limit (see (\ref{rho_limit})), $\rho$ is given by two- and three-point correlation functions of $\phi_{1,3}$ operators.  For spin (resp.\ FK) clusters at $Q=3$, $h_{1,3}=7/5$ (resp.\ $2/3$), while for $Q=2$, $h_{1,3}=5/3$ (resp.\ $1/2$).  Thus the correlation functions decay more rapidly with distance in the spin case, making the events  ${\bar P}_{\ell}$, ${\bar P}_r$, and ${\bar P}_{\ell,r}$  less likely than the events  $P_{\ell}$, $P_r$, and $P_{\ell,r}$ in the FK case, so that simulation errors increase.   Assuming that the ratio of the deviation for the  spin case to the FK case is roughly equal to corresponding ratio of $\rho$ values, we find (for $x$ near $R/2$) a ratio of order 10 in either case which is consistent with our results.   

  Figures (\ref{spinQ3}) and (\ref{Rhalf}) shows that the $Q=3$ case ($\kappa=10/3$) also exhibits these features.  The most noticeable difference between the $Q=2$ and $Q=3$ cases is the considerably larger error in the latter when $R=3$.  Because the system size of the latter is quadruple that of the former, the $Q=3$ run time for a single sample is quadruple the $Q=2$ single sample run time.  Also because more samples need to be generated in order to achieve a reasonable accuracy, we were unable to achieve the same degree of accuracy as with the $Q=2$ case.  The $Q=3$ data fluctuates about the theory curve, so the large error is most likely attributable to noise that can be reduced with more samples.

\subsection{Density of percolation crossing clusters with free boundary conditions} \label{Dfbc}

The density $P_{A_x}$ of horizontal percolation crossing clusters with free boundary conditions on all sides of $\mathcal{R}$ is given by (\ref{xingdens}).  The  data (resp.\ prediction) was normalized by dividing by the measured (resp.\ predicted) value  in the center of the rectangle.   In order to simulate $P_{A_x}$ for percolation, we carried out a simulation similar to what was done in \cite{SimmonsZiffKleban08}, in which we ``grew" clusters from all the sites at one edge of the rectangle.  For those clusters that reached the opposite edge, we added all wetted sites to an array that tallied the number of times each site of the lattice was visited.  Dividing by the total number of runs gave the density.  If one cluster crossed, we did not include the sites of any other, non-crossing clusters, as would be the case for fixed (``wired") boundary conditions.    We considered square
and rectangular lattices, results for  $128 \times 128$ and $128 \times 256$ are shown.  For larger aspect ratios, the statistical significance dropped because of the lower probability of finding a crossing cluster. For each, we carried out about $10^8$ samples.

In figure \ref{PiAxplots} we show a comparison between the theory and the measured values.  The contour lines are at heights $0.5, \, 0.6, \,0.7, \,0.8, \,0.87, \,0.93, \,0.96,\,0.98, \,0.994, \,0.999$, from the outside inwards.  The deviations are quite small.  For instance, for $R=2$ and various fixed $x$ values, we found the following relative errors (absolute value of the difference between predicted and measured value divided by the predicted value) averaged along the $y$ coordinate: ($x,\,{\rm error})=$ $(1,\,0.0094)$, $(0.5,\,0.0057)$, $(0.25,\,0.0038)$, $(0.125,\,0.0036)$.  And for various fixed $y$ values: ($y,\,{\rm error})=$ $(0.5,\,0.011)$, $(0.25,\,0.0055)$, $(0.125,\,0.0030)$, $(0.0625,\,0.0021)$.

\section{Summary  \label{Summ}}

In this section we summarize the main results in some detail, and reproduce the main formulas.  The research done here exploits,   for a variety of two-dimensional critical models, the solutions for a chiral six-point correlation function in order to specify  the density of critical clusters  anchored to one or both vertical sides of a rectangle with wired boundary conditions on those sides and free boundary conditions on the horizontal sides. These models include the critical $\mathrm{O}(n)$ loop models in both high and low density phases and both Fortuin-Kasteleyn (FK) and spin clusters in the critical $Q$-state Potts models. We also determine the density of percolation crossing  clusters (from one side of a rectangle to the opposite) with free boundary conditions.

Subsection \ref{cfDEs} reviews the solutions  found in \cite{SimmonsKleban11}  for the correlation function  (\ref{cf1})
\be \nn
C(z)=\langle \phi_{1,2}^c(0)\phi_{1,2}^c(\ci) \Phi_{1/2,0}(z, \bar z) \phi_{1,2}^c(R)\phi_{1,2}^c(R+\ci) \rangle_{\cal R}\; ,
\ee 
in the rectangular geometry  $\mathcal{R}:= \{z=x + \ci y\in\mathbb{C}\, |\, 0<x<R, 0<y<1\}$.  Here $z=x+iy$, $C$ also depends on the parameter $m$ that specifies the aspect ratio of the rectangle (see (\ref{mvsR})), and the solutions hold for arbitrary SLE parameter $\k > 0$. They appear in the form (\ref{CeqfG})
\be \nn
C(z) = f(\xi,\psi,m) \, G\left(\xi,m \right) \; ,
\ee
where the algebraic prefactor $f$ is given by (\ref{fxipsi}) and (\ref{fconst}). It is independent of boundary conditions, and in fact a function of $x$ and $y$ via the coordinates $\xi = \xi(x,m)$ and $\y=\y(y,m)$,   given by elliptic functions (see (\ref{xipsidef})).  $\xi$ and $\y$ are the natural coordinates for this problem, as explained in  \cite{SimmonsKleban11}.  The factor $G$ is a single conformal block, given by an algebraic factor and an Appell hypergeometric function  (see (\ref{GI}-\ref{GV}) and (\ref{GVI})).  $G$ depends on boundary conditions, and determining how the various $G$s contribute to the cluster densities for the different boundary conditions of interest is one of our main results. ($G$ can depend either on $\xi$ and $m$, as written here, or alternatively on $\y$ and $m$ as mentioned in  subsection \ref{cfDEs}, depending, respectively, on whether the right and left or top and bottom sides of the rectangle are `wired.')

Subsections \ref{pie} and \ref{Icc} next examine  the question of expressing  cluster densities (resp.\ crossing probabilities) for the boundary conditions of interest (see figure \ref{config3}) in terms of the $G$s (resp.\ hypergeometric functions), and also explains how the various cases are implemented in specific models.  Making use of results in \cite{BauerBernardKytola05}, an explicit form for a horizontal crossing probability that generalizing Cardy's  horizontal crossing probability for percolation to these models is given in (\ref{Hxings}) (see also the lines just below this equation). Expressions for the density $P_\ell(z,R)$ (resp.\ $P_r(z,R),P_{\ell r}(z,R)$) of clusters anchored to the left (resp.\ right, left and right) side(s) of a rectangle with either independently  wired left and right sides in terms of the weights $\Pi_A$, $\Pi_{B_\ell}$, $\Pi_{B_r}$, $\Pi_H$ and $\Pi_V$ (and the corresponding densities ${\bar P}_\ell(z,R)$ (resp.\ ${\bar P}_r(z,R),{\bar P}_{\ell r}(z,R)$) for mutually wired sides) are given in (\ref{Pllr})-(\ref{Plrlr}). (Explicit formulas for $\Pi_H$ and $\Pi_V$ are found in (\ref{Cardy11}), (\ref{Cardy13}), (\ref{G13}) and (\ref{PiVeq}).)  Subsection \ref{Icc} then employs Coulomb gas methods to find solutions for the weights $\Pi_A$, $\Pi_{B_\ell}$ and $\Pi_{B_r}$ (the first row of figure \ref{config3}). These  are given in the three equations (\ref{pia})-(\ref{pib2}),
\bea \nn
\Pi_\AT  &=& f(\xi,\psi,m)\, G_{\rm VI}(\xi,m)\\ \nn
\Pi_{\BR}&=& f(\xi,\psi,m)\, G_{\rm I}(\xi,m) \qquad\qquad \mathrm{and}\\ \nn
\Pi_{\BL}&=& f(\xi,\psi,m)\, G_{\rm V}(\xi,m)\; .
\eea
The expressions for the bottom row of figure \ref{config3} follow by symmetry.  This completes our determination of the densities for the models and boundary conditions of interest here.

Section \ref{factor} then discusses the factorization behavior implied by our solutions, which generalizes previous results for percolation  to a variety of critical models.  For percolation, the universal ratio (\ref{rho6})
\be \nn
\rho_{\mathrm{perc}}(x, R) =  \frac{P_{\ell r}(z,R)}{\sqrt{P_r(z,R) P_{\ell}(z,R) P_\Ho(R)}}\;, 
\ee
was considered \cite{SimmonsZiffKleban08}, with $P_\Ho(R)$ the horizontal crossing probability given by Cardy. 
In that case, $\rho$ is constant to within $3 \%$ inside the rectangle, being equal to $1$ when $z$ is on the left or right side, and rising to about $1.03$ when $z$ is far from the sides.  The numerator, a six-point correlation function, therefore  factorizes into the lower-order correlations in the denominator to very good approximation. In this section, we generalize $\rho$  to an expression (\ref{rho}) defined in terms of correlation functions restricted to be single conformal blocks by specification of the propagating channels. This then leads to the expression for $\rho$  (\ref{rho3}) in terms of the weights $\Pi_A$, $\Pi_{B_\ell}$, $\Pi_{B_r}$, $\Pi_H$ and $\Pi_V$ or equivalently the conformal blocks $G$ and the quantities $G_{1,1}$ and $G_{1,3}$, valid for all the critical models mentioned.   In all cases, by definition $\rho=1$ when $z$ is on the left or right side. It is also constant when $z$ it is far from the sides, but how much that constant deviates from $1$ depends on the model (more exactly, on $\k$).  Thus the factorization changes significantly with $x$ in some models.   
 However, in all cases $\rho$ is independent of $y$, as in percolation,  because the algebraic prefactor $f$, which also depends on $y$ via $\y$, divides out of the ratio, leaving only conformal blocks $G$ and other factors that are independent of $y$.  This $y$-independence is a consequence of the unusual symmetry of the conformal blocks $G$ found in \cite{SimmonsKleban11}. 

Next, section \ref{Dpcc} makes use of the locality property of percolation to derive the explicit result (\ref{xingdens}) giving the density of percolation clusters in a rectangle with free boundary conditions on all sides
\begin{align} \nn
P_{\AX}&= 2^{1/3} (K')^{5/48}\left[\frac{(1-m \xi^2)^2}{ \xi (1-\xi) (1-m \xi)} + \frac{(1-(1-m) \psi^2)^2}{ \psi (1-\psi) (1-(1-m) \psi)} -4 \right]^{-11/96}   \\ \nn
&\hspace{1in}
\Bigg[ 
\left(\frac{1-m\, \xi}{1-\xi}+ m\frac{1- \xi}{1-m\, \xi}  -\frac{(1-(1-m)\y)^2+m}{1-(1-m)\y}\right)^{1/6}+\left(\frac{1}{\xi}+m\, \xi-\frac{(1-(1-m)\y)^2+m}{1-(1-m)\y}\right)^{1/6}\\ \nn
&\hspace{1in}
-\frac{(1-m)^{1/3 }}{\left[\xi (1-\xi )(1-m \xi )\right]^{1/6}}F_1\left(\frac{1}{3};\frac{2}{3},-\frac{2}{3};\frac{2}{3}\bigg| m,m \xi \right) \Bigg]\; , 
\end{align}
with $F_1$ the Appell function (\ref{F1def}) and $K'(m)=K(1-m)$, with $K$ the complete elliptic integral.

In section \ref{Sims} we compare our predictions for the ratio $\rho$ (see (\ref{rho3})), which governs factorization, for the $Q=1$, $2$ and $3$-state Potts models, including both FK and spin clusters, for various aspect ratios.   The agreement is very good.  The formula (\ref{xingdens}) for the density of crossing clusters in percolation is also verified.

The Appendix presents forms for the densities and crossing weights  in models where the hypergeometric functions simplify.

\section{Acknowledgments}

JJHS is grateful for useful conversations with I. Gruzberg.
 
 This work was supported by  National Science Foundation Grants Nos.\ MRSEC DMR-0820054 (JJHS),  DMR-0536927 (PK and SMF), PHY-0855335 (SMF) and DMS-0553487 (RMZ).


\appendix{}

%
%
\section{Special parameter values} \label{Spv}

In this appendix we consider special values of the parameter $\k$ for which any of the hypergeometric series that occur in the   density conformal blocks or crossing weights (see    (\ref{GI}-\ref{GV}) and (\ref{Cardy11}-\ref{Cardy13}) respectively)   may be written in terms of simpler functions.  In particular we look at $\k$ values for which the functions have non-positive integer parameters, as this may imply that the series truncates into hypergeometric series in fewer variables, potentially even into a polynomial.    These five quantities are either proportional to the Appell function $ F_1$, which appears  in the forms
\begin{align} \label{A1}
u_1(s,t) &:= F_1\left(1-\frac{4}{\k};\frac{4}{\k},2-\frac{16}{\k};2-\frac{8}{\k}\bigg|s,t \right)\\ \label{A2}
u_2(s,t)&:=\frac{\G(2-8/\k)\G(16/\k -1)}{\G(12/\k)\G(1-4/\k)}F_1\left(1-\frac{4}{\k };\frac{4}{\k },\frac{4}{\k };\frac{12}{\k }\bigg| s,t \right)\; ,
\end{align}
in (\ref{GII},\, \ref{GIII},\, \ref{GIV}) and  (\ref{GI},\, \ref{GV}) 
respectively; or as one of the functions $G_{1,1}(m)$ and $G_{1,3}(m)$, defined in (\ref{Cardy11},\, \ref{Cardy13}), which are given by a hypergeometric function ${}_2F_1$.
In what follows we consider values $2\le\k\le8$ for which either $2-8/\k$, $2-12/\k $ or $2-16/\k \in \mathbb{Z}\le0$.  There are ten such values: $\k \in \{8,\, 6,\, 16/3,\, 4,\, 16/5,\,  3,\, 8/3,\, 16/7,\, 12/5,\, 2 \}$, which correspond to $n \in \{0,\, \pm1,\, \pm\sqrt{2},\, \pm2 \}$ in the critical $\mathrm{O}(n)$ loop model dilute phase and $n \in \{0,\, 1,\, \sqrt{2}\}$ in the dense phase.

The common pre-factor given in (\ref{fxipsi}) is already an algebraic function in $\xi$ and $\psi$.    Thus, for brevity we will not always include it, and   restrict our attention to the functions (\ref{GI}-\ref{GVI}) and (\ref{Cardy11},\, \ref{Cardy13}).

We also note that in some cases hypergeometric series with rational parameters of small denominator can be written in terms of other relatively simple functions.  The ten $\k$ values that we discuss are all those with $2\le\k\le8$ that  give rational hypergeometric parameters with denominator of four or less.  Because the denominator value has a pronounced effect on the type of functions that appear upon simplifying the hypergeometric series, and because the associated values of $n$ for each denominator of four or less all have the same magnitude, we group these values together.

\subsection[Integer  Parameters]{Integer hypergeometric parameters: $|n|=2$}
\subsubsection{Dilute $n=2$: Gaussian Free Field}

Parameter value $\k=4$ corresponds to the critical Gaussian free field or to the $Q=4$ state Potts model. For this value the various hypergeometric functions are
\begin{equation*}
\begin{array}{r@{\;}l@{\qquad}r@{\;}l}
u_1(s,t)&={\displaystyle \frac{2-s-t(2-t)}{2(1-s)}}&
u_2(s,t)&={\displaystyle\frac{1}{2}}\\[2ex]
G_{1,1}(m)& ={\displaystyle\frac{2-m}{2}}&
G_{1,3}(m)& ={\displaystyle\frac{m}{2}}\; .
\end{array}
\end{equation*}
The conformal blocks are proportional to
\begin{equation*}
\begin{array}{r@{\;}l@{\hspace{0.75cm}}r@{\;}l}
G_{\rm I}(\xi,m)&=\frac{\displaystyle m(1-m)\xi ^2}{\displaystyle\sqrt{4m(1-m)\xi(1-\xi)(1-m \xi)}}&
G_{\rm II}(\xi,m)&=\frac{\displaystyle (1-m)(1+m \xi^2)}{\displaystyle\sqrt{4m(1-m) \xi (1-\xi )(1-m \xi )}}\\
G_{\rm III}(\xi,m)&=\frac{\displaystyle 2-m-2m \xi+m^2 \xi^2}{\displaystyle\sqrt{4m(1-m) \xi (1-\xi )(1-m \xi )}}&
G_{\rm IV}(\xi,m)&=\frac{\displaystyle (1-m \xi )^2+m(1-\xi)^2}{\displaystyle\sqrt{4m(1-m)\xi (1-\xi )(1-m \xi)}}\\
G_{\rm V}(\xi,m)&=\frac{\displaystyle m(1-\xi )^2}{\displaystyle\sqrt{4m(1-m)\xi(1-\xi)(1-m \xi )}}&
G_{\rm VI}(\xi,m)&=\frac{\displaystyle (1-m)(1-m \xi^2)}{\displaystyle\sqrt{4m(1-m)\xi(1-\xi)(1-m \xi)}}
=\sqrt{\displaystyle\frac{1-m}{\displaystyle m\, {\rm sn}^2(2x K'|m)}} \; .
\end{array}
\end{equation*}

   The expression (\ref{rho3}) for $\rho$ becomes 
\be
\rho(x)=\sqrt{\frac{1+m}{1+m\, \je{cn^2}{2x}}}\; ,
\ee
where    double argument identities for the Jacobi elliptic functions have been used.

\subsubsection{Dilute $n=-2$: Loop-Erased Random Walk}

An SLE with parameter $\k=2$ has been shown to correspond to the loop erased random walk \cite{Schramm2000_UST_LERW}. However, the loop fugacity is less than zero for that model,  so the present correlation function does not translate to a set of densities.  As $\k \to 2$ the hypergeometric functions become
\begin{equation*}
\begin{array}{rl@{\qquad}rl}
u_1(s,t) &={\displaystyle \frac{1+s-3t}{2}+\frac{(1-t)^5}{2(1-s)^3}(1-2s+2 t-s t)}&
u_2(s,t) &={\displaystyle\frac{s+t-3}{2}}\\[2ex]
G_{1,1}(m) & ={\displaystyle 1-2m+m^3-\frac{m^4}{2}}&
G_{1,3}(m) & ={\displaystyle -m^3+\frac{m^4}{2}}\; .
\end{array}
\end{equation*}
And we identify a few key functions:
\begin{align*}
G_{\rm I}(\xi,m)&=-\frac{m(1-m)\xi^{5}\left[3-\xi-m\xi\right]}{2\left[ \xi(1-\xi)(1-m\, \xi)\right]^{3/2}}\\
G_{\rm V}(\xi,m)&=-\frac{m(1-\xi)^{5}\left[2-m+\xi-2 m \xi\right]}{2(1-m)^2\left[ \xi(1-\xi)(1-m\, \xi)\right]^{3/2}}\\
G_{\rm VI}(\xi,m)&=-\frac{(1-m)\left[1+m-3m \xi +3m^3\xi ^5-m^3\xi ^6-m^4\xi ^6\right]}{2m^2\left[ \xi(1-\xi)(1-m\, \xi)\right]^{3/2}}=-\frac{(1-m)\left[4+4m-3m{\rm sn}^2(2K' x|m)\right]}{m^2 {\rm sn}^3(2K' x|m)}
\end{align*}

\subsection[Half integer parameters]{Half integer hypergeometric parameters: $n=0$}

For fugacity $n=0$ in either the dense ($\k=8$) or dilute ($\k=8/3$) phase $\rho(x,R)=1$, as discussed.  A peculiar property of $n=0$ models is the absence of bulk loops.    Therefore every point (except for those that belong to the SLE hulls) is adjacent to some part of the boundary, so that the bulk density operator is a generalization of the zero weight indicator operator used by Schramm in \cite{SchrammPercForm01}.

Schramm's paper considered a single SLE hull in a simply connected domain, and calculated the probability that the SLE passed to the left of a marked point in the bulk.  This is equivalent to a correlation function with two SLE operators to generate the SLE hull, and a bulk indicator at the marked point that returns one if the hull passes to its left and $0$ otherwise.  If it is possible (not possible) to draw a path from the point to the right boundary without crossing the hull the hull passes left (right) of the marked point.  Thus, when $n=0$ and the only loop in the configuration is the hull itself the interpretation of the indicator and the density operator are identical, as evidenced by the weight $2 h_{1/2,0}=0$, which is also the the weight of Schramm's operator.

In the current correlation function, the bulk operator acts as an indicator operator in the presence of the two SLE hulls generated by the four $\phi_{1,2}$ operators.  The indicator distinguishes between six cases: three correspond to the possible locations of a point relative to a horizontal crossing of hulls (configurations $\AT$, $\BB$, and $\BT$ in figure \ref{config3}), and three correspond to the analogous case for vertical crossing (configurations $\AV$, $\BL$, and $\BR$).

The sum of the   weights   with horizontal crossings equals the total weight of horizontal crossings.  A similar result holds for vertical crossings. These relationships can be written as
\begin {align} \label{Equ: PH_Cond}
\Pi_\Ho&\propto \Pi_\AT+\Pi_\BT+\Pi_\BB\\ \label{Equ: PV_Cond}
\Pi_\Ve&\propto \Pi_\AV+\Pi_\BR+\Pi_\BL\; .
\end{align}
The   weights   from the bottom row of figure \ref{config3} follow from (\ref{pia}-\ref{pib2}) upon letting $(\xi,m,\AT,\BR,\BL) \to (\y,1-m,\AV,\BT,\BB)$. To emphasize the relations between the densities and crossing weights, we include the pre-factor $(\ref{fxipsi})$ in this section.

\subsubsection{Dense $n=0$: Peano Curve}
We first examine the dense phase with $\k=8$. This case is equivalent to a space-filling Peano curve.  The relevant hypergeometric functions reduce to elliptic integrals on applying well-known identities:
\begin{align*}
u_1(x,y) &=G_{1,1}(x)=G_{1,3}(x)=
{}_2 F_1\left(\frac{1}{2},\frac{1}{2};1\bigg|x \right)=\frac{2}{\pi}K(x)\\
u_2(x,y)&=\frac{2}{\pi}F_1\left(\frac{1}{2};\frac{1}{2},\frac{1}{2};\frac{3}{2}\bigg| x,y \right)=\frac{2}{\pi \sqrt{x}}\,  \mathrm{F}\left(\sin^{-1}\sqrt{x}\bigg|\frac{y}{x}\right)\; ,
\end{align*}
where $\mathrm{F}(\cdot|m)$ is the incomplete elliptic integral of the first kind.
This satisfies ${\rm F} \left( \sin^{-1} {\rm sn}(z|m) |m\right)=z$, i.e.\ it acts as an inverse elliptic function.

For $\k=8$ the function $f_{\k=8}(\xi,\psi,m)=(K')^{-1}[m(1-m)]^{-1/4}$ and we find:
\begin{equation*}
\begin{array}{r@{\:}l@{\hspace{1.5cm}}r@{\:}l}
\Pi_\Ho&
{\displaystyle =\frac{ 2 K' }{\pi K'}=\frac{2}{\pi}}&
\Pi_\BR&
{\displaystyle =\frac{2}{\pi K'} {\rm F}\left(\sin^{-1}\sqrt{\xi} \Big| m\right)=\frac{2 x}{\pi}} \\[2ex]
\Pi_\Ve&
{\displaystyle =\frac{2 K }{\pi K'}=\frac{2 R}{\pi}}&
\Pi_\BL&
{\displaystyle =\frac{2}{\pi K'}{\rm F}\left(\sin^{-1}\sqrt{\frac{1-\xi}{1-m\, \xi}}\,\Bigg| m\right)=\frac{2(R-x)}{\pi}}\\[2ex]
\Pi_\AT=\Pi_\BT+\Pi_\BB&
{\displaystyle=\frac{2 K' }{\pi K'}=\frac{2}{\pi}}&
\Pi_\BT&
{\displaystyle=\frac{2}{\pi K'} {\rm F}\left(\sin^{-1}\sqrt{\psi} \Big|1-m\right)=\frac{2 y}{\pi}} \\[2ex]
\Pi_\AV=\Pi_\BR+\Pi_\BL&
{\displaystyle=\frac{2 K}{\pi K'}=\frac{2 R}{\pi}}&
\Pi_\BB&
{\displaystyle=\frac{2}{\pi K'}{\rm F}\left(\sin^{-1}\sqrt{\frac{1-\psi}{1-(1-m)\, \psi}}\,\Bigg| 1-m\right)=\frac{2(1-y)}{\pi}}\; .
\end{array}
\end{equation*}
We simplify these expressions using elliptic function identities that are related to the symmetry operations in (\ref{Sym}).

The value $\k=8$ corresponds to a space filling Peano curve, equivalent to the the boundary between the uniform spanning tree (UST) \cite{Schramm2000_UST_LERW} anchored to the vertical sides and its dual, which is another UST anchored to the horizontal sides.  As the curve is space filling, any infinitesimal neighborhood intersects both the UST and its dual, so the densities of spanning trees and dual spanning trees are equal and uniform throughout the rectangle.

Only one of the UST and its dual can be a single component tree, since by connecting the left and right edges we prevent the connection of the top and bottom and vice versa.  Thus when the UST spans the rectangle from left to right there are two components to the dual clusters, one attached to the top edge and the other to the bottom.   Along with the uniform density of the trees, this means that $\Pi_\Ho\propto\Pi_\AT=\Pi_\BT+\Pi_\BB$,    the same as   \eqref{Equ: PH_Cond}.  A similar result holds when the left and right sides belong to different trees, which implies a vertical crossing by the dual tree.\

\subsubsection{Dilute $n=0$: Self Avoiding Walks}
The value of $\k=8/3$ corresponds to the problem of self avoiding walks.  For $\k=8/3+\epsilon$ the hypergeometric expressions  all diverge with $\Gamma(2-8/\k) \sim -8/9\epsilon$, as $\epsilon \to 0$.  We adjust the normalization of $u_1, u_2, G_{1,1}$ and $G_{1,3}$, dividing by a factor of $-\G(2-8/\k)$ to eliminate this divergence:
\begin{align*}
\widetilde{G}_{1,1}(m):=\frac{G_{1,1}(m)}{-\G(2-8/\k)}&=\frac{ 15}{32}m^2(1-m)^2 {}_2F_1\left(\frac{7}{2},\frac{3}{2};3\bigg|m\right) =\left[\frac{1-m+m^2}{\pi}E(m)-\frac{2-3m+m^2}{2\pi}K(m)\right] \\
\widetilde{G}_{1,3}(m):=\frac{G_{1,3}(m)}{-\G(2-8/\k)}&=\widetilde{G}_{1,1}(m)\\
\tilde{u}_1(s,t):=\frac{u_1(s,t)}{-\G(2-8/\k)} 
&=\left[\frac{(s-2s t+t^2)^2}{s^2(1-s)^2}-\frac{8t(s-t)(1-t)}{5s(1-s)}\frac{\partial}{\partial s}\right]\widetilde{G}_{1,1}(s) \; ,
\end{align*}
   where $E(m)$ is the complete elliptic integral of the second kind. Thus similar to  $\k=8$, the gaussian hypergeometric functions are replaced by complete elliptic integrals.

The crossing weights are
\begin{align*}
\Pi_\Ho(m)&=(K')^5 \left[\frac{1-m+m^2}{\pi}E(1-m)-\frac{m(1+m)}{2\pi}K(m)\right]\\
\Pi_\Ve&=(K')^5 \left[\frac{1-m+m^2}{\pi}E(m)-\frac{2-3m+m^2}{2\pi}K(m)\right]\; .
\end{align*}

The density pre-factor is
$$
f_{\k=8/3}(x,y,m)=\frac{(K')^5 \left[m(1-m)\right]^{5/4}}{{\rm ds}^2(2K' x|m)+{\rm ds}^2(2K' y|1-m)},
$$
which, along with the simplified expressions for the various $G(\xi,m)$ functions, gives us
\begin{align*}
\Pi_\BR+\Pi_\BL&=\frac{4{\rm ds}^2(2K' x|m)\Pi_\Ve-(8/5)m (1-m) \partial_m \Pi_\Ve}{{\rm ds}^2(2K' x|m)+{\rm ds}^2(2K' y|1-m)}\\
\Pi_\AT&=\frac{4{\rm ds}^2(2K' x|m)\Pi_\Ho-(8/5)m(1-m) \partial_m \Pi_\Ho}{{\rm ds}^2(2K' x|m)+{\rm ds}^2(2K' y|1-m)}\\
\Pi_\BT+\Pi_\BB&=\frac{4{\rm ds}^2(2K' y|1-m)\Pi_\Ho+(8/5)m (1-m) \partial_m \Pi_\Ho}{{\rm ds}^2(2K' x|m)+{\rm ds}^2(2K' y|1-m)}\\
\Pi_\AV&=\frac{4{\rm ds}^2(2K' y|1-m)\Pi_\Ve+(8/5)m(1-m) \partial_m \Pi_\Ve}{{\rm ds}^2(2K' x|m)+{\rm ds}^2(2K' y|1-m)}\; ,
\end{align*}
after simplifying with double argument identities for the elliptic functions.
With the densities written in this way, relations (\ref{Equ: PH_Cond}) and (\ref{Equ: PV_Cond}) follow immediately.

\subsection[Parameters with denominator three]{Hypergeometric parameters with denominator three: $|n|=1$}

For these particular parameters ($\k=12/5$, $3$ or $6$) the hypergeometric functions do not simplify beyond the truncation of $G_{1,1}(m)$ that occurs because $0 \ge 2-12/\k \in \mathbb{Z}$.

The truncated functions, which are related to the identity channel of the crossing weights (see (\ref{G11})) are,
\be \nonumber
G_{1,1}(m)= \left\{
\begin{array}{l@{\qquad}rl}
1&\k=&6\\
1-m+m^2&\k=&3\\
(1-2m)(1+m)(1-m/2)&\k=&12/5
\end{array}\right. \;.
\ee

The SLE parameter $\k=6$ corresponds to critical percolation, or equivalently the dense phase of the ${\rm O}(1)$ loop gas.  This is a purely probabilistic limit. Here the form of the identity channel crossing weight is such that the sum of all crossing probabilities is one. 

The SLE parameter $\k=3$ corresponds to spin clusters in the critical Ising model, or equivalently the dilute phase of the ${\rm O}(1)$ loop gas.  This is proportional to the Ising partition function in a rectangle with the  horizontal and vertical edges fixed in opposite spins.

The SLE parameter $\k=12/5$ corresponds to the dilute phase of the ${\rm O}(-1)$ loop gas.

\subsection[Parameters with denominator four]{Hypergeometric parameters with denominator four: $|n|=\sqrt{2}$}

For $\k=16/3$ or $16/5$, which correspond to $n^2=Q=2$, our expressions  simplify greatly.

\subsubsection{Dense $n=\sqrt{2}$: Ising FK clusters}
An SLE with parameter $\k=16/3$ corresponds to the boundary of the critical $Q=2$ FK cluster model.  For $\k = 16/3$ the hypergeometric functions become
\begin{equation*}
\begin{array}{rl@{\qquad}rl}
G_{1,1}(m) & ={\displaystyle \sqrt{\frac{1+\sqrt{1-m}}{2}}}&
G_{1,3}(m) & ={\displaystyle \sqrt{\frac{1-\sqrt{1-m}}{2}}}\\[2ex]
u_1(s,t) &={\displaystyle \sqrt{\frac{1+\sqrt{1-s}}{2(1-s)}}-t\sqrt{\frac{1-\sqrt{1-s}}{2s(1-s)}}}%
\; .
\end{array}
\end{equation*}

The functions that simplify become
\begin{align*}
G_{\rm II}
&=\frac{(1-m)^{3/8}}{\sqrt{2}m^{1/8}\left[\xi(1-\xi)(1-m \xi)\right]^{1/4}}
\frac{1+\sqrt{m}\xi}{\sqrt{1+\sqrt{m}}}\\
G_{\rm IV}
&=
\frac{(1-m)^{3/8}}{\sqrt{2}m^{1/8}\left[\xi(1-\xi)(1-m \xi)\right]^{1/4}}
\frac{1-\sqrt{m}\xi}{\sqrt{1-\sqrt{m}}}\\
G_{\rm III}&=\left[G_{\rm II}+G_{\rm IV} \right]/\sqrt 2\; .
\end{align*}
Combining these with
$$
G_{\rm VI}
=\frac{(1-m)^{3/8}}{\sqrt{2}m^{1/8}\left[\xi(1-\xi)(1-m \xi)\right]^{1/4}}
\left[\frac{1+\sqrt{m}\xi}{\sqrt{1+\sqrt{m}}}-\frac{32 \sqrt{m \pi} \xi^{5/4}}{5\G(1/4)^2}F_1\left(\frac{1}{4};\frac{3}{4},\frac{3}{4};\frac{9}{4}\bigg|\xi,m \xi \right)\right]\; ,
$$
gives the relatively simple expression 
\be
\rho(x)=\sqrt{\frac{1+\sqrt{m}\xi}{1-\sqrt{m}\xi}}-\frac{32\xi^{5/4}}{5\G(1/4)^2}\sqrt{\frac{m(1+\sqrt{m}) \pi}{1-m \xi^2}}F_1\left(\frac{1}{4};\frac{3}{4},\frac{3}{4};\frac{9}{4}\bigg|\xi,m \xi \right) \; .
\ee

\subsubsection{Dilute $n=\sqrt{2}$}

An SLE with parameter $\k=16/5$ corresponds to the dilute phase of the ${\rm O}\left(\sqrt{2}\right)$ loop gas.  At $\k = 16/5$ the hypergeometric functions become
\begin{align*}
u_1(s,t)
&=\frac{1}{4}\left[
2\frac{s^{3/2}-t^3}{s^{3/2}}\sqrt{1-\sqrt{s}}
-\frac{\left(t-\sqrt{s}\right)^3}{s \left(1-\sqrt{s}\right)^{3/2}}
+2\frac{s^{3/2}+t^3}{s^{3/2}}\sqrt{1+\sqrt{s}}
-\frac{\left(t+\sqrt{s}\right)^3}{s \left(1+\sqrt{s}\right)^{3/2}}\right]\\
G_{1,1}(m)
&=\left[2-3\sqrt{1-m}+2(1-m)\right]\left(\frac{1+\sqrt{1-m}}{2}\right)^{3/2}\\
G_{1,3}(m)
&=\left[2+3\sqrt{1-m}+2(1-m)\right]\left(\frac{1-\sqrt{1-m}}{2}\right)^{3/2}\; .
\end{align*}
This means that
\begin{align}
G_{\rm II}
&=\frac{\left(1-\sqrt{m}\right)^{3/2}\left[ 2\left(1+\sqrt{m}\right)^2 \left(1+m^{3/2} \xi^3\right)- \sqrt{m} \left(1+\sqrt{m}\xi \right)^3 \right]}
{16\, m^{7/8}\left[ \xi(1-\xi)(1-m\, \xi)\right]^{3/4}\left(1-m\right)^{7/8}}\\
G_{\rm IV}
&=\frac{\left(1+\sqrt{m}\right)^{3/2}\left[ 2\left(1-\sqrt{m}\right)^2 \left(1-m^{3/2} \xi^3\right)+ \sqrt{m} \left(1-\sqrt{m}\xi \right)^3 \right]}
{16\, m^{7/8}\left[ \xi(1-\xi)(1-m\, \xi)\right]^{3/4}\left(1-m\right)^{7/8}}\\
G_{\rm III}&=\left[G_{\rm II}+G_{\rm IV} \right]/\sqrt 2 \; .
\end{align}

\subsubsection{Dilute $n=-\sqrt{2}$}

The final case corresponds to $\k=16/7$, the dilute phase of the ${\rm O}(-\sqrt{2})$ loop gas.
At $\k = 16/7$ the hypergeometric functions become
\begin{align*}
G_{1,1}(m)&=
-\left(\frac{m}{2}\right)^{5/2}\frac{4 (1-m)^2 - 10 (1-m)^{3/2}+11 (1-m) - 10 (1-m)^{1/2}+4}{\left(1- \sqrt {1-m}\right)^{5/2}}\\
G_{1,3}(m)&=
-\left(\frac{m}{2}\right)^{5/2}\frac{4 (1-m)^2 +10 (1-m)^{3/2}+11 (1-m)+10 (1-m)^{1/2}+4}{\left(1+\sqrt {1-m}\right)^{5/2}}\\
u_1(s,t)
&=\sum_{i=0}^5 \binom{5}{i}(-t)^i\,{}_2F_1\left(-3/4+i,7/4;-3/2+i| s\right)\\
&=\left[ \frac{1-5t+10t^2}{(1-s)^{5/2}}-\frac{20t(1-4t)}{13(1-s)^{3/2}}\frac{\partial}{\partial s}+\frac{160 t^2}{117(1-s)^{1/2}}\frac{\partial^2}{\partial s^2} \right]G_{1,1}(s)-\\
&\qquad
\left[ \frac{t^3(t^2-5st+10s^2)}{[s(1-s)]^{5/2}}-\frac{20 t^3(t-4s)}{13[s(1-s)]^{3/2}}\frac{\partial}{\partial s}+\frac{160 t^3}{117[s(1-s)]^{1/2}}\frac{\partial^2}{\partial s^2} \right]G_{1,3}(s)\; .
\end{align*}
In this case,     the expressions being rather lengthy, we leave it to the interested reader to assemble the simplified functions and densities.

\bibliography{RedDensRZ}

\begin{thebibliography}{10}

\bibitem{SimmonsKleban11}
Jacob J.~H. Simmons and Peter Kleban.
\newblock Complete conformal field theory solution of a chiral six-point
  correlation function.
\newblock {\em ArXiv e-prints}, 1103.2458, 2011.

\bibitem{LanglandsEtAl92}
R.~P. Langlands, C.~Pichet, Ph. Pouliot, and Y.~Saint-Aubin.
\newblock On the universality of crossing probabilities in two-dimensional
  percolation.
\newblock {\em J. Stat. Phys.}, 67:553--574, 1992.

\bibitem{Cardy92}
John~L. Cardy.
\newblock Critical percolation in finite geometries.
\newblock {\em J. Phys. A: Math. Gen.}, 25(4):L201--L206, 1992.

\bibitem{Smirnov}
S.~Smirnov.
\newblock Critical percolation in the plane.
\newblock {\em C. R. Acad. Sci. Paris Sr. I Math.}, 333:239, 2001.

\bibitem{MathieuRidout07}
P.~{Mathieu} and D.~{Ridout}.
\newblock {From percolation to logarithmic conformal field theory}.
\newblock {\em Phys. Lett. B}, 657:120--129, 2007.

\bibitem{Kytola08}
K.~{Kyt{\"o}l{\"a}}.
\newblock From {SLE} to the operator content of percolation.
\newblock {\em ArXiv e-prints}, 0804.2612, 2008.

\bibitem{SimmonsKlebanZiffJPA07}
Jacob J.~H. Simmons, Peter Kleban, and Robert~M Ziff.
\newblock Percolation crossing formulae and conformal field theory.
\newblock {\em J. Phys. A: Math. Th.}, 40(31):F771--F784, 2007.

\bibitem{LawlerSchrammWerner01}
Gregory Lawler, Oded Schramm, and Wendelin Werner.
\newblock Values of {B}rownian intersection exponents, {I}: Half-plane
  exponents.
\newblock {\em Acta Math.}, 187:237--273, 2001.

\bibitem{KlebanZagier03}
Peter Kleban and Don Zagier.
\newblock Crossing probabilities and modular forms.
\newblock {\em J. Stat. Phys.}, 113:431--454, 2003.

\bibitem{DiamantisKleban09}
Nikolaos Diamantis and Peter Kleban.
\newblock New percolation crossing formulas and second-order modular forms.
\newblock {\em Communications in Number Theory and Physics}, 3:677--696, 2009.

\bibitem{KlebanSimmonsZiff06}
Peter Kleban, Jacob J.~H. Simmons, and Robert~M. Ziff.
\newblock Anchored critical percolation clusters and 2d electrostatics.
\newblock {\em Phys. Rev. Lett.}, 97(11):115702, 2006.

\bibitem{SimmonsKlebanZiffPRE07}
Jacob J.~H. Simmons, Peter Kleban, and Robert~M. Ziff.
\newblock Exact factorization of correlation functions in two-dimensional
  critical percolation.
\newblock {\em Phys. Rev. E}, 76(4):041106, 2007.

\bibitem{SimmonsKleban09}
Jacob J.~H. Simmons and Peter Kleban.
\newblock {General solution of an exact correlation function factorization in
  conformal field theory.}
\newblock {\em J. Stat. Mech: Th. Exp.}, 2009:P10002, 2009.

\bibitem{SimmonsZiffKleban08}
Jacob J.~H. Simmons, Robert~M. Ziff, and Peter Kleban.
\newblock Factorization of percolation density correlation functions for
  clusters touching the sides of a rectangle.
\newblock {\em J. Stat. Mech: Th. Exp.}, 2009:P02067, 2009.

\bibitem{NienhuisRiedelSchick_1980}
B.~Nienhuis, E.~K. Riedel, and M.~Schick.
\newblock Magnetic exponents of the two-dimensional $q$-state {Potts} model.
\newblock {\em J. Phys. A: Math. Gen.}, 13(6):L189, 1980.

\bibitem{FortuinKasteleyn1972}
C.~M. Fortuin and P.~W. Kasteleyn.
\newblock On the random-cluster model: {I.} {I}ntroduction and relation to
  other models.
\newblock {\em Physica}, 57(4):536--564, 1972.

\bibitem{sw}
R.~Swendsen and J.~Wang.
\newblock Nonuniversal critical dynamics in {M}onte {C}arlo simulations.
\newblock {\em Phys. Rev. Lett.}, 58:86--88, 1987.

\bibitem{RohdeSchramm2005}
{S.~Rohde} and {O. Schramm}.
\newblock Basic properties of {SLE}.
\newblock {\em Ann. Math.}, {\bf 161}:883--924, 2005.

\bibitem{ChelkakSmirnov2009}
{D.~Chelkak} and {S. Smirnov}.
\newblock Universality in the 2d {Ising} model and conformal invariance of
  fermionic observables.
\newblock {\em ArXiv eprints}, 0910.2045, 2009.

\bibitem{GamsaCardy2007}
Adam Gamsa and John Cardy.
\newblock Schramm-{L}oewner evolution in the three-state {P}otts model: A
  numerical study.
\newblock {\em J. Stat. Mech: Th. Exp.}, 2007(08):P08020, 2007.

\bibitem{BauerBernardKytola05}
M.~{Bauer}, D.~{Bernard}, and K.~{Kyt{\"o}l{\"a}}.
\newblock Multiple {Schramm-Loewner} evolutions and statistical mechanics
  martingales.
\newblock {\em J. Stat. Phys.}, 120:1125, 2005.

\bibitem{ArguinSaintAubin2002}
Louis-Pierre Arguin and Yvan Saint-Aubin.
\newblock Non-unitary observables in the 2d critical {Ising} model.
\newblock {\em Phys. Lett. B}, 541:384--389, 2002.

\bibitem{DotsenkoFateev84}
Vl.~S. Dotsenko and V.~A. Fateev.
\newblock {Conformal algebra and multipoint correlation functions in 2d
  statistical models.}
\newblock {\em Nucl. Phys. B}, 240:312--348, 1984.

\bibitem{BYB}
P.~{DiFrancesco}, P.~{Mathieu}, and D.~{S{\' e}n{\' e}chal}.
\newblock {\em Conformal Field Theory}.
\newblock Springer, 1999.

\bibitem{Cardy84}
J.~L. Cardy.
\newblock Conformal invariance and surface critical behavior.
\newblock {\em Nucl. Phys.}, {\bf B240}:514--532, 1984.

\bibitem{SimmonsKleban07}
Jacob J.~H. Simmons and Peter Kleban.
\newblock {First column boundary operator product expansion coefficients.}
\newblock {\em ArXiv e-prints}, 0712.3575v1, 2007.

\bibitem{Schramm2000_UST_LERW}
Oded Schramm.
\newblock Scaling limits of loop-erased random walks and uniform spanning
  trees.
\newblock {\em Israel J. Math.}, 118:221--288, 2000.

\bibitem{SchrammPercForm01}
O.~Schramm.
\newblock A percolation formula.
\newblock {\em Electronic Comm. Prob.}, 6:115--120, 2001.

\end{thebibliography}
\end{document}